\documentclass[fleqn,11pt]{wlscirep}

\usepackage[utf8]{inputenc}
\usepackage[T1]{fontenc}
\usepackage[nolist,nohyperlinks]{acronym}

\usepackage{indentfirst}

\title{Multiscale Modeling of Vacancy-Cluster Interactions and Solute Clustering Kinetics in Multicomponent Alloys}
\author[1]{Zhucong Xi}
\author[2]{Louis G. Hector Jr.}
\author[1,3]{Amit Misra}
\author[1,*]{Liang Qi}
\affil[1]{Department of Materials Science and Engineering, University of Michigan, Ann Arbor, Michigan 48109, USA}

\affil[2]{GM Global Technical Center, General Motors Company, Warren, Michigan 48092, USA}

\affil[3]{Department of Mechanical Engineering, University of Michigan, Ann Arbor, Michigan 48109, USA}

\affil[*]{qiliang@umich.edu}
\begin{abstract}
Prediction of solute clustering kinetics in aged multicomponent alloys requires a quantitative understanding of complex vacancy-cluster interactions across multiple scales. Here, we develop an integrated computational framework combining on-lattice kinetic Monte Carlo (KMC) simulations, absorbing Markov chain models, and mesoscale cluster dynamics (CD) to investigate these interactions in Al-Mg-Zn alloys. The Markov chain model yields vacancy escape times from solute clusters and identifies a two-stage behavior of the vacancy-cluster binding energy. These binding energies are used to estimate residual vacancy concentrations in the Al matrix after quenching, which serve as critical inputs to CD simulations to predict long-term cluster evolution kinetics during natural aging. Our results quantitatively demonstrate the significant impact of quench rate on natural aging kinetics. Results provide insights to guide alloy chemistry, quench rates, and aging time at finite temperature to control the evolution of solute clusters and eventual precipitates in aged multicomponent alloys.
\end{abstract}

\begin{document}
\begin{acronym}
    \acro{CMC}{canonical Monte Carlo}
    \acro{CNT}{classical nucleation theory}
    \acro{CD}{cluster dynamics}
    \acro{DFT}{density functional theory}
    \acro{FCC}{face-centered cubic}
    \acro{GP}{Guinier-Preston}
    \acro{GB}{grain boundary}
    \acrodefplural{GB}{grain boundaries}
    \acro{KMC}{kinetic Monte Carlo}
    \acro{MC}{Monte Carlo}
    \acro{PEL}{potential energy landscape}
    \acro{PF}{phase-field}
    \acro{SA}{simulated annealing}
    \acro{SQS}{special quasirandom structures}
\end{acronym}

\flushbottom
\maketitle

\thispagestyle{empty}

\section*{Introduction}
Age-hardenable alloys, such as 6000 and 7000 series aluminum (Al) alloys, derive their mechanical strength from the controlled formation of solute clusters and precipitates through appropriate heat treatments that activate vacancy-mediated diffusion. Rapid quenching after solution treatment introduces a high concentration of residual vacancies, which significantly accelerates early-stage precipitation kinetics by enhancing the mobility of substitutional solute atoms~\cite{raabe2022making}. At typical solutionizing temperatures ($\sim$500~$^\circ$C), the equilibrium vacancy concentration in Al is approximately $\sim10^{-4}$, but it drops to around $\sim10^{-11}$ at room temperature~\cite{azarniya2019recent,carling2003vacancy}. Achieving this equilibrium requires extended timescales due to slow annihilation of vacancies at dislocations and \acp{GB}. As a result, vacancy supersaturation can persist for hours (or even days) after quenching~\cite{wang2024modelling,chen2025using}. Beyond their basic role in solute diffusion, vacancies interact strongly with solute atoms and clusters~\cite{shin2010firstprinciples,peng2020solutevacancy,chen2023investigation,wolverton2007solute,wang2025revisiting}, dynamically altering the population of mobile vacancies. For instance, trace Sn additions in Al-Mg-Si alloys can effectively trap vacancies and suppress the formation of Mg-Si clusters during natural aging~\cite{pogatscher2014diffusion,werinos2016design}. Moreover, growing solute clusters themselves can bind vacancies, which reduces the available mobile vacancy concentration and slows further clustering kinetics~\cite{vincent2006solute,vincent2008precipitation,soisson2007cuprecipitation,jain2023natural}. Therefore, a quantitative understanding of vacancy-cluster interactions and their impact on vacancy evolution is essential for accurately modeling and controlling diffusion-driven precipitation processes in age-hardenable alloys~\cite{borgenstam2000dictra,dumitraschkewitz2019sizedependenta}, which ultimately determine key mechanical properties such as strength and ductility.

These complex interactions between vacancies and growing solute clusters in multicomponent alloys remain relatively underexplored by computational methods due to their chemical complexity and the inherently multi-time- and multi-length-scale nature of the underlying diffusion processes. Atomistic methods, such as on-lattice \ac{KMC} simulations, can accurately model vacancy migration kinetics using local-environment-dependent migration barriers informed by first-principles calculations~\cite{mantina2009first,vanderven2018firstprinciples,osetsky2016specific}. However, these methods face inherent limitations in both spatial and temporal scales. Spatial constraints arise from the need to explicitly track individual atomic and vacancy jumps~\cite{leetmaa2014kmclib,mitchell2023parallel}, while temporal limitations stem from frequent low-barrier events that drastically reduce simulation efficiency~\cite{chatterjee2010accurate,puchala2010energy,fichthorn2013local}. Furthermore, conventional on-lattice \ac{KMC} approaches struggle to capture the dynamic evolution of mobile vacancy concentrations~\cite{vincent2006solute,vincent2008precipitation,soisson2007cuprecipitation}, which is essential for predicting long-term diffusion and phase transformation behavior. To address these challenges, computational methods, such as \ac{PF} simulations~\cite{li1998computer,vaithyanathan2002multiscale,kleiven2020precipitate}, are needed to capture the feedback between evolving vacancy concentrations and solute clustering on the mesoscale. However, these mesoscale models require inputs such as mobility functions and initial cluster configurations, which ideally should be derived from atomistic simulations based on actual alloy compositions and heat treatment histories. Yet, this connection remains difficult due to the aforementioned limitations of atomistic methods. Consequently, a significant knowledge gap persists in linking atomic-scale vacancy-mediated diffusion with mesoscale phase transformation modeling.

Bridging the knowledge gap in vacancy-mediated diffusion processes is particularly critical for 7000 series Al-Mg-Zn alloys. These alloys can achieve high strengths (up to approximately 700~MPa ultimate tensile strength) after appropriate heat treatments~\cite{liddicoat2010nanostructural}. However, their broader application, such as in the automotive industry, is limited because the rapid formation of solute clusters during natural aging significantly reduces ductility and formability~\cite{sha2004earlystage, liu2015effect, huo2016warm}. Our recent studies~\cite{chatterjee2022situ, xi2024kinetic} revealed that Mg-Zn-rich solute clusters form rapidly during the quenching process, rather than only after reaching room temperature. On-lattice \ac{KMC} simulations show that these early-formed clusters strongly trap vacancies, leading to a significant reduction in the mobile vacancy concentration and potentially slowing subsequent precipitation kinetics~\cite{xi2024kinetic}. This phenomenon, often referred to as ``vacancy prisons''~\cite{jain2021machine, pogatscher2011mechanisms}, corresponds to deep energy basins on the \ac{PEL} that govern vacancy migration~\cite{jain2023natural}. Importantly, however, vacancies trapped in these clusters are not completely immobile. Vacancy migration barriers within Mg-Zn-rich solute clusters are generally lower than in the surrounding Al matrix, allowing relatively fast vacancy diffusion within the cluster region and enabling a finite probability for vacancies to escape back into the matrix~\cite{xi2024kinetic}.

Although these qualitative descriptions provide useful insight into vacancy trapping, the overall effects on long-time solute clustering kinetics are still unclear due to the limitations of atomistic scale methods described above. A quantitative understanding of vacancy-cluster interactions remains limited. Zurob and Seyedrezai~\cite{zurob2009model} proposed a model wherein the vacancy trapping effect is proportional to the cluster size (i.e., the number of solute atoms), which has been validated for small clusters in Al alloys~\cite{lay2012vacancy,pogatscher2012interdependent}. This relates to the growing spatial extent of the energy basin. In contrast, Soisson \textit{et al.}~\cite{soisson2007cuprecipitation} introduced a different approach for Cu clusters in the Fe-Cu system, emphasizing that the binding energy is governed by the difference in vacancy formation energies between the Cu precipitate and the Fe matrix,  which dictates the energetic depth of the trapping basin. While these models capture different aspects of the \ac{PEL}, their application to multicomponent systems, such as Al-Mg-Zn, is hindered by the combinatorial complexity of cluster configurations and the complex interactions among different solute species. As a result, accurately modeling vacancy-cluster interactions in such systems remains a significant challenge, and conventional approaches (e.g., simple binding energy estimates) may not sufficiently account for this complexity~\cite{myhr2024modeling,jain2023natural}.

Fortunately, similar challenges involving low-energy basins have been extensively studied in the context of accelerating \ac{KMC} simulations~\cite{novotny1995montea,athenes1997identification,athenes2000effects,deo2002firsta,mason2004stochastic,daniels2020hybrid,dokukin2018efficient,athenes2019elastodiffusion}. Conventional \ac{KMC} approaches often become inefficient when systems are repeatedly trapped in states separated by small energy barriers, leading to prolonged residence in low-energy basins. To address this issue, several advanced methods have been developed. In particular, Chatterjee and Voter~\cite{chatterjee2010accurate}, Puchala \textit{et al.}~\cite{puchala2010energy}, and Fichthorn and Lin~\cite{fichthorn2013local} introduced absorbing Markov chain frameworks that accelerate \ac{KMC} simulations by efficiently treating transitions within such basins. Building upon these strategies, we apply an absorbing Markov chain model~\cite{puchala2010energy,fichthorn2013local} to systematically quantify the low-barrier, low-energy basins associated with Mg-Zn-rich solute clusters in Al-Mg-Zn alloys. In this approach, solute clusters are modeled as bidirected graphs, where nodes represent possible vacancy positions and edges denote transitions between nearest-neighbor lattice sites. By analyzing the transition probabilities and absorbing state behavior, we compute vacancy escape times from clusters and validate our results against direct \ac{KMC} simulations. Additionally, we evaluate how effective vacancy binding energies vary with cluster size and assess their influence on the concentration of residual mobile vacancies in the Al matrix, a critical factor governing long-time diffusion kinetics after quenching. Our Markov chain model reveals a two-stage behavior in vacancy-cluster binding energies: an initial size-dependent regime for small clusters, consistent with the model proposed by Zurob and Seyedrezai~\cite{zurob2009model}, followed by a saturation regime governed by the vacancy formation energy difference between the cluster and matrix, in line with the findings of Soisson \textit{et al.}~\cite{soisson2007cuprecipitation}. Further details are provided in the Results section.

Finally, by leveraging the effective vacancy binding energies as functions of cluster size, which is quantified through the integrated \ac{KMC} and absorbing Markov chain framework described above, we obtain essential input parameters for mesoscale simulations to capture long-time diffusion kinetics and associated diffusional phase transformations. \Ac{CD} simulations provide a mesoscale modeling approach to describe the temporal evolution of atomic clusters (e.g., solute-vacancy clusters, precipitates, or defect aggregates) using population-based rate equations, without explicitly tracking individual atoms~\cite{clouet2005precipitation,clouet2009modeling,lepinoux2009modelling,lepinoux2010comparing,lepinoux2018multiscale}. These rate equations incorporate the vacancy-cluster binding energy functions derived from our atomistic framework. Accordingly, we construct \ac{CD} models to investigate long-time clustering kinetics, allowing mobile vacancy concentrations to evolve dynamically with the solute cluster configurations. The predicted results correspond well to experimentally relevant time and length scales. Our findings quantitatively demonstrate that cooling rate strongly influences natural aging kinetics, offering predictive insights for optimizing heat treatments in engineering alloys. Additionally, the predicted cluster distributions provide critical initial conditions for solute nucleation and early-stage growth in \ac{PF} simulations of precipitation kinetics~\cite{li1998computer,vaithyanathan2002multiscale,kleiven2020precipitate}.

\section*{Results}
\subsection*{Kinetic Monte Carlo simulations during quenching}
\begin{figure}[!ht]
    \centering
    \includegraphics[width=0.8\linewidth]{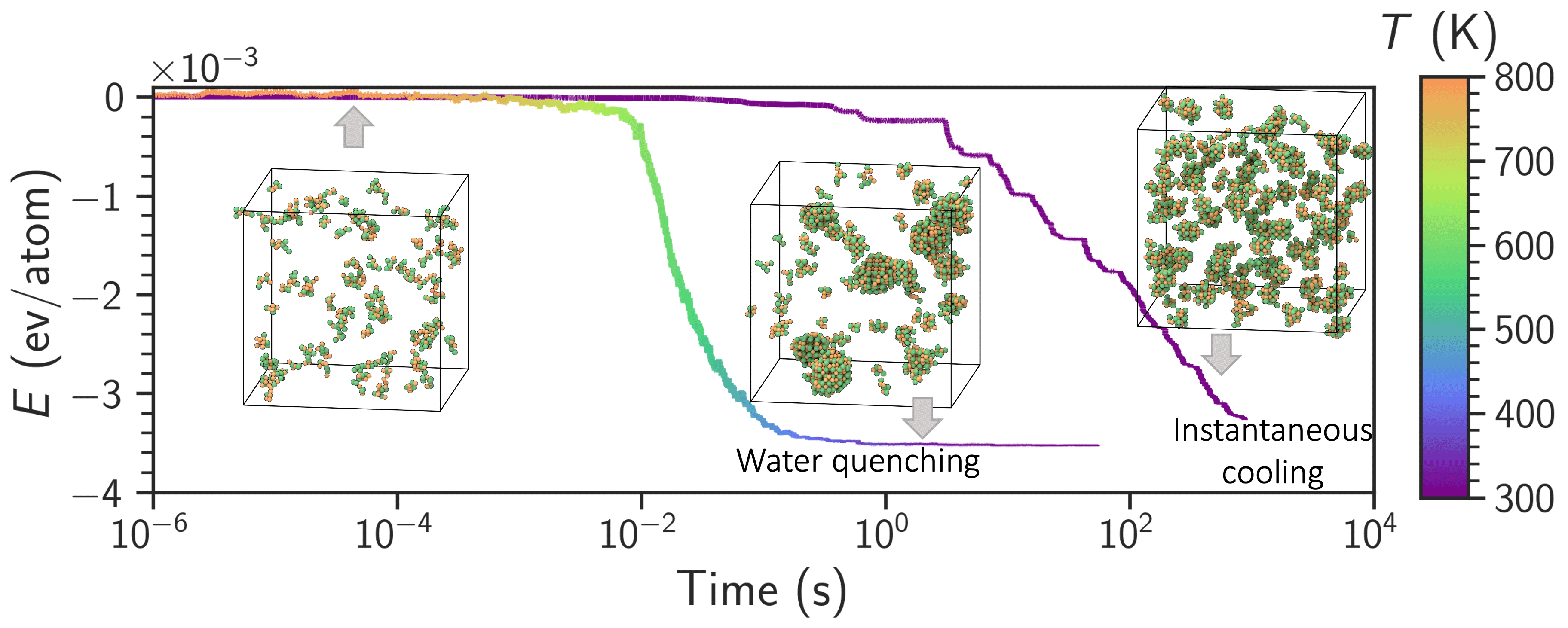}
    \caption{Time evolution of the average energy per atom $E$ of two typical KMC quenching simulations\cite{xi2024kinetic}. The curve above is at a constant temperature of 300~K from the first step, representing an infinitely fast (instantaneous) cooling simulation from a solid solution. The curve below is from a water-quenching simulation. The change in the simulation temperature is denoted by the colors defined in the right color contour bar. Snapshots illustrate the configurations before and after quenching from these simulations. The length of the cubic simulation supercells in the simulation snapshots is 12 nm. Clusters here are defined as solute atoms (Mg and Zn) that are connected as their 1$^{\text{st}}$ nearest neighbors. Al atoms surrounded by 12 clustered solute atoms as their first nearest neighbors are also counted. In the snapshots, only clusters with more than 10 atoms are presented.} %Illustration of `vacancy prison' generated by solute clusters from KMC simulations.
    \label{fig:trapping_in_kmc}
\end{figure}

The details of the on-lattice \ac{KMC} simulation methods are provided in our recent studies and briefly summarized in the Methods Section~\cite{xi2024kinetic}. To illustrate the vacancy trapping effects due to solute clusters, Fig.~\ref{fig:trapping_in_kmc} presents time evolutions of the average energy per atom $E$ from two \ac{KMC} quenching simulations of an Al-2.86 at.$\%$Mg-2.38 at.$\%$Zn alloy (within the composition of 7075 Al alloy). Both simulations start with the same initial configuration of a random solid solution state, which is obtained by annealing at 800 K, a typical solutionizing temperature for Al alloys~\cite{azarniya2019recent}, using the \ac{CMC} method~\cite{xi2024kinetic}. The purple curve in Fig.~\ref{fig:trapping_in_kmc} represents the case where the sample is instantaneously quenched from the solid solution state at 800 K to 300 K and held at 300 K during natural aging. In contrast, the color-gradient curve in Fig.~\ref{fig:trapping_in_kmc} represents a water-quenching scenario, where the temperature gradually decreases from 800 K to 300 K, as indicated by the color of the curve. Although Mg-Zn-rich solute clusters form in both cases, as shown in the inset subfigures, the corresponding energy evolution curves exhibit markedly different behaviors. In the instantaneous cooling case, the energy begins to gradually decrease approximately $10^{-1}$ second after cooling is initiated and continues to decline even beyond $10^3$ seconds. In contrast, the energy in the water quenching case drops sharply between $\sim10^{-2}$ and $\sim10^{-1}$ seconds, during which the temperature falls from approximately 600~K to 400~K. After this interval, the energy stabilizes as the temperature continues to decrease below 400~K, eventually reaching $\sim$300~K. These results indicate that solute clustering kinetics are most active within the intermediate temperature range of approximately 400-600~K but are significantly suppressed below $\sim$400~K. 

The distinct energy evolution profiles in Fig.~\ref{fig:trapping_in_kmc} also lead to different solute cluster configurations, as shown in the inset figures visualized by OVITO\cite{stukowski2010visualization}: the instantaneous cooling case results in a higher density of smaller clusters, whereas the water quenching case produces fewer but larger clusters. Another significant difference between these two cases lies in the solute clustering kinetics at low temperatures ($\sim$300~K). In the instantaneous cooling case, the system's energy continues to decrease at 300~K, indicating ongoing, albeit slow, vacancy-mediated solute diffusion. In contrast, energy evolution ceases entirely in the water quenching case, suggesting that vacancies become effectively trapped by solute clusters formed during quenching~\cite {xi2024kinetic}. 

\subsection*{Energetic analysis of vacancy-cluster interactions}
To reveal the mechanism behind vacancy trapping in Al-Mg-Zn alloys, we first aimed to determine the structures and energetics of the clusters. Although \ac{DFT} calculations offer reliable energetic estimations~\cite{xi2022mechanism}, identifying the most stable structures of clusters with the given numbers of Mg ($n_{\text{Mg}}$) and Zn ($n_{\text{Zn}}$) atoms in Al-Mg-Zn multicomponent alloys remains challenging because the possible configurations of solute clusters explode exponentially with the increase of $n_{\text{Mg}}$ and $n_{\text{Zn}}$. Hence, we employed surrogate models trained on first-principles data to compute the energy of a supercell with a specified lattice occupation configuration (see Supplementary Note 1~\cite{xi2022mechanism}). A \ac{SA} heuristic~\cite{kirkpatrick1983optimization}, adapted from the Metropolis-Hastings algorithm~\cite{metropolis1953equationa}, was then applied to generate clusters (see Supplementary Note 2). This method efficiently explores configuration space while minimizing the average formation enthalpy per solute atom.
%, denoted as $\bar{H}_\text{form}$

\begin{figure}[!htb]
    \centering
    \includegraphics[width=0.8\linewidth]{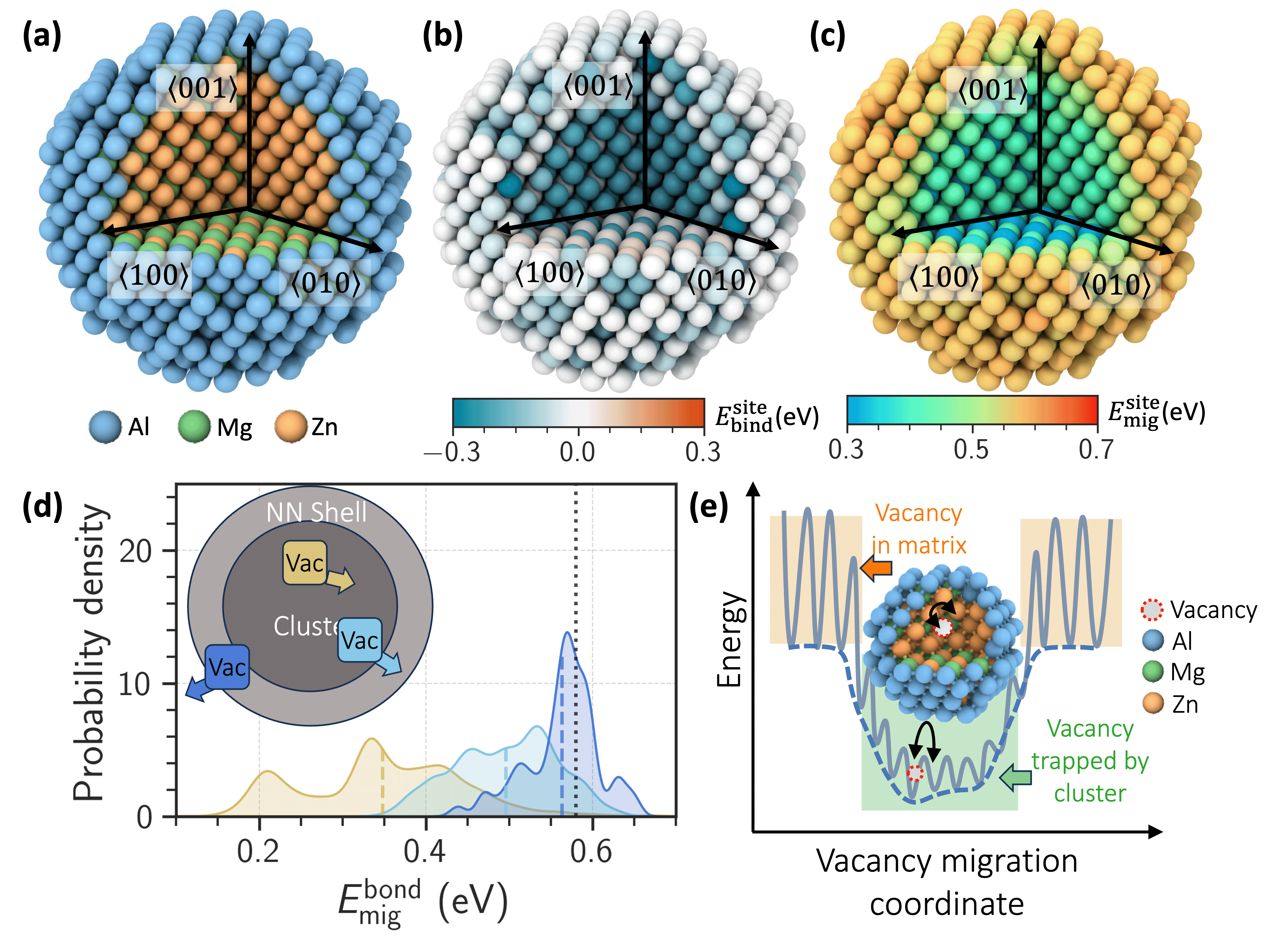}
    \caption{Analyses of local cluster energetics from SA simulations based on surrogate models trained on first-principles data~\cite{xi2022mechanism,xi2024kinetic}. (a)-(c) An example solute cluster showing its internal structures: (a) spatial distribution of elemental species, (b) local vacancy binding energy at each site, $E^{\text{site}}_{\text{bind}}$, and (c) local average vacancy migration barrier at each site, $E^{\text{site}}_{\text{mig}}$. (d) Probability density plots obtained using kernel density estimation (KDE) for the local vacancy migration barrier associated with each first-nearest-neighbor bond, $E^{\text{bond}}_{\text{mig}}$, in this example cluster. Three distributions are shown, corresponding to different migration event types defined in the inset schematic: yellow indicates migration events within the cluster interior; light blue denotes vacancy jumps from the cluster surface to its first nearest-neighbor (NN) shell; dark blue represents vacancy jumps from the NN shell to the surrounding matrix. Dashed colored lines indicate average values for each event type, and the black dotted line marks the migration barrier in pure Al, $E_{\text{mig}}^{\text{Al}} = 0.58$~eV. (e) Schematic illustration of a conceptual ``vacancy prison'' in the potential energy landscape (PEL) of vacancy migration around a solute cluster. The dashed line traces local energy minima, while the solid curved line connects both minima and transition states. Curved double-headed arrows represent vacancy migration. }
    \label{fig:cluster_energies}
\end{figure}

After the structures of the clusters were obtained, we examined the local vacancy binding energy at each lattice site, $E^{\text{site}}_{\text{bind}}$, and the local vacancy migration barriers associated with each atomic bond, $E^{\text{bond}}_{\text{mig}}$, between two nearest-neighbor sites. Figs.~\ref{fig:cluster_energies}a-c depict an example cluster extracted from the \ac{SA} simulations, sliced to show its internal structure. Fig.~\ref{fig:cluster_energies}a highlights the atomic arrangement of Al (blue), Mg (green), and Zn (orange). This cluster exhibits a structure similar to the reported \ac{GP}-I zones~\cite{lervik2021atomic}, where the number of Mg-Zn bonds is maximized, although it lacks strict periodicity. Fig.~\ref{fig:cluster_energies}b presents the spatial distribution of $E_{\text{bind}}^{\text{site}}$ calculated from the following expression: 
\begin{equation}
\label{eq:site_bind_energy}
E_{\text{bind}}^{\text{site}} = 
\left[E\left(\text{Vac}\right) - E\left({\text{Al}}_{N-1}\text{Vac}\right) \right] - \left[E\left(\text{X}\right)- E\left({\text{Al}}_{N-1}\text{X}\right)\right],
\end{equation}
where $E(\text{Vac})$ denotes the energy of the supercell containing a vacancy at a particular site in the cluster, and $E(\text{X})$ represents the energy of the same supercell with the atom X to fill up the previous local vacancy site. Atom X is the element originally present at that site, which can be Al, Mg, or Zn. Similarly, $E({\text{Al}}_{N-1}\text{Vac})$ and $E({\text{Al}}_{N-1}\text{X})$ correspond to the energies of a larger, pure Al supercell with a vacancy and with atom X occupying the vacancy site, respectively. This equation compares the formation energy difference of a vacancy-free cluster to the same cluster but containing a vacancy at a particular site. The value of $E_{\text{bind}}^{\text{site}}$ quantifies how strongly a vacancy binds at a particular site relative to the solute that occupies the position. Specifically, in a dilute Al matrix, $E_{\text{bind}}^{\text{site}} = 0$, indicating no additional binding for the vacancy. In contrast, within a cluster, $E_{\text{bind}}^{\text{site}}$ typically exhibits a negative value, suggesting that vacancies are more likely to be bound or trapped within the cluster compared to the Al matrix. 

Fig.~\ref{fig:cluster_energies}c similarly presents the spatial distribution of the local average vacancy migration barriers at each site, $E_{\text{mig}}^{\text{site}}$. This value represents the arithmetic average of the 12 vacancy migration barriers, $E_{\text{mig}}^{\text{bond}}$, associated with atomic bonds connecting each site to its 12 first-nearest neighbors, assuming the site is initially occupied by a vacancy. The results highlight numerous sites within the cluster that exhibit low average migration barriers, with values around $\sim 0.3$ eV. These values of $E_{\text{mig}}^{\text{site}}$ are notably lower than those at sites near the cluster shell, which present higher values similar to the vacancy migration barrier in pure Al, $E_\text{mig}^{\text{Al}}$, of $0.58$ eV.

To further quantify the variations in migration barriers among different site types, Fig.~\ref{fig:cluster_energies}d presents the probability distributions of $E_{\text{mig}}^{\text{bond}}$ for three categories of vacancy migration events: (i) migrations within the cluster (yellow), (ii) transitions from the cluster surface to the surrounding nearest-neighbor shell (light blue), and (iii) jumps beyond the nearest-neighbor shell (dark blue). Consistent with Fig.~\ref{fig:cluster_energies}c, migration barriers are generally lower within the cluster interior. In contrast, vacancy jumps beyond the nearest-neighbor shell predominantly exhibit barriers in the range of 0.54 to 0.62~eV, closely aligning with the migration barrier in pure Al. Finally, Fig.~\ref{fig:cluster_energies}e provides a conceptual illustration of the \ac{PEL} of vacancy migration, representing the so-called ``vacancy prison'' created by a solute cluster. An energy basin is spatially localized at the cluster scale, with its boundary effectively defined by the cluster-matrix interface, where $E_{\text{bind}}^{\text{site}} \approx 0$ and $E_{\text{mig}}^{\text{bond}} \approx E_{\text{mig}}^{\text{Al}}$. The energy basin strongly binds vacancies, significantly reducing their mobility and effectively trapping them. Although trapped, vacancies within these "prisons" are not completely immobile. They can still move around within the cluster due to relatively low migration barriers. However, their probability of escaping back into the matrix is greatly reduced, leading to prolonged trapping.

\subsection*{Absorbing Markov chain model for vacancy escape from clusters}

\begin{figure}[!htbp]
    \centering
    \includegraphics[width=0.8\linewidth]{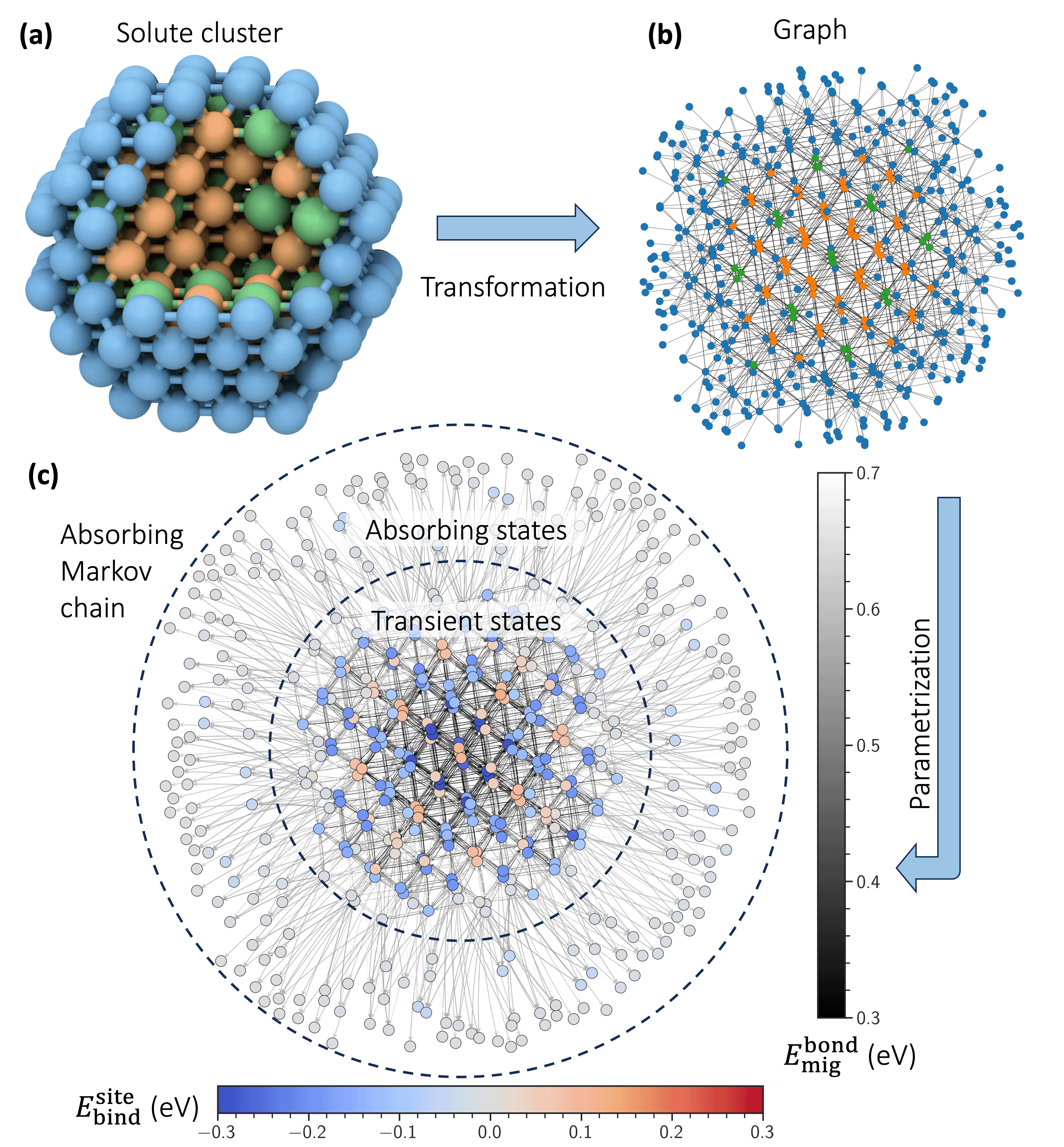}
    \caption{Workflow for constructing an absorbing Markov chain model to analyze vacancy-cluster interactions and predict vacancy escape times. (a) Atomic configuration of an example solute cluster, showing atomic arrangements and elemental distributions, with colors indicating Al (blue), Mg (green), and Zn (orange). (b) Graph representation of the cluster, including the cluster atoms and an additional layer of neighboring atoms. Nodes correspond to lattice sites occupied by atoms or vacancies, and edges represent bonds between nearest neighbors. (c) Absorbing Markov chain (bidirected graph) constructed from the graph in (b), where nodes are colored by local vacancy binding energies, $E_{\text{bind}}^{\text{site}}$, and edges are shaded according to the migration barriers between connected sites, $E_{\text{mig}}^{\text{bond}}$. The outermost layer denotes absorbing states, representing sites where vacancies are considered to have escaped from the cluster region.}
    \label{fig:flowchart}
\end{figure}

We construct a numerical model to quantitatively describe the vacancy escape process from the ``vacancy prison'' illustrated in Fig.~\ref{fig:cluster_energies}d based on an absorbing Markov chain framework (see Methods section for the related concepts), using the energetics obtained from the surrogate model (Figs.~\ref{fig:cluster_energies}b-c)~\cite{xi2022mechanism,xi2024kinetic}. The workflow used to build the absorbing Markov chain model, predict vacancy escape times, and analyze vacancy-cluster interaction energetics is illustrated in Fig.~\ref{fig:flowchart}. As depicted in Fig.~\ref{fig:flowchart}a, we first obtained solute cluster structures directly from \ac{SA}, although clusters can similarly be derived from \ac{KMC} or other atomistic simulation methods. Then, the atomic structure is transformed into a graph representation (Fig.~\ref{fig:flowchart}b). Each atom is represented as a graph node, and each atomic bond as a graph edge. To accurately describe the event vacancy jumping out of the nearest neighbor shell (Fig.~\ref{fig:cluster_energies}d), an additional layer of neighboring atoms surrounding the cluster is included. Next, the graph is parameterized to become a bidirected graph, as shown in Fig.~\ref{fig:flowchart}c. This transformation introduces directionality into the graph, labeling the information of vacancy migration paths. The outermost layer in this bidirected graph represents absorbing states, which are viewed as states in the Al matrix, where vacancies are considered to have escaped from the cluster region. The local binding energy, $E_{\text{bind}}^{\text{site}}$, is calculated for each node using the surrogate model to characterize the thermodynamic stability of vacancies at specific lattice sites. Similarly, vacancy migration barriers between connected nodes, $E_{\text{mig}}^{\text{bond}}$, are also computed using the surrogate model to parameterize transition probabilities between different sites.

Nodes (lattice sites) in the cluster and the nearest neighbor shell (lattice sites as the nearest neighbors of any sites in the cluster) are treated as transient states, whereas nodes (lattice sites) in the outermost layer (off the nearest neighbor shell) act as absorbing states. Transient states represent sites that vacancies can temporarily occupy and subsequently migrate away from, whereas absorbing states denote sites that are considered as the vacancy having escaped the cluster. Fig.~\ref{fig:flowchart}c presents the same information as Figs.~\ref{fig:cluster_energies}b-c: interior nodes generally exhibit lower (more negative) binding energies, signifying stronger vacancy binding, while bonds inside the cluster typically show lower migration barriers, highlighting easier vacancy migration within the cluster.

Finally, the absorbing Markov chain model is constructed based on the graph representation of Fig.~\ref{fig:flowchart}c. The mean exit time that a vacancy escapes from the cluster, $t_{\text{esc}}$, can be determined using the formula~\cite{fichthorn2013local}:
\begin{equation}
\label{eq:exit_time}
t_{\text{esc}} = \vec{p_0}^T (\pmb I - \pmb T)^{-1} \vec{\tau},
\end{equation}
Here $\pmb{T}$ is the transient matrix containing transition probabilities among all transient states, which is a square matrix of size $n_\text{tr} \times n_\text{tr}$, where $n_\text{tr}$ is the number of transient states and determined by sites in the cluster and the nearest neighbors. $\pmb I$ is an identity matrix that has the same size as $\pmb{T}$. $\vec{p_0}$ is a vector to represent the initial probabilities of transient states with a size of $n_\text{tr} \times 1$, and $\vec{\tau}$ is a vector of one-step escape times for each site, which has the same size as $\vec{p_0}$. 

Specifically, elements $T_{ij}$ from the matrix $\pmb{T}$ are derived from the migration barriers $E^{\text{bond}}_{\text{mig}}$ based on an Arrhenius relationship:
\begin{equation}
\label{eq:T}
T_{ij} = \frac{\exp\left(-\frac{E^{\text{bond}}_{\text{mig}, ij}}{k_B T}\right)}{\sum_{j=1}^{z} \exp\left(-\frac{E^{\text{bond}}_{\text{mig}, ij}}{k_B T}\right)},
\end{equation}
where $z$ is the coordination number ($z=12$ for an \ac{FCC} lattice), $E^{\text{bond}}_{\text{mig}, ij}$ is the migration barrier from node $i$ to node $j$, corresponding to one of the 12 vacancy migration barriers from a lattice site to a nearest neighbor site. To obtain the values of elements $p_i$ in $\vec{p_0}$, an equilibration assumption has been employed~\cite{puchala2010energy, fichthorn2013local, daniels2020hybrid}. It assumes that each transient state will be visited numerous times since the energy basin cannot be escaped easily, eventually reaching a local equilibrium within the basin. Once this local equilibrium is established, the probability of occupying any given transient state $i$ follows a Boltzmann distribution:
\begin{equation}
p_i = \frac{\exp\left({-\frac{E^{\text{site}}_{\text{bind},i}}{k_B T}}\right)}{\sum_{j=1}^{n_\text{tr}}\exp\left({-\frac{E^{\text{site}}_{\text{bind},i}}{k_B T}}\right)},
\label{eq:p_0}
\end{equation}
where $E^{\text{site}}_{\text{bind},i}$ is the energy of vacancy on node $i$. The denominator sums contributions from all $n_{\text{tr}}$ transient states. The element, $\tau_i$, in the vector $\vec{\tau}$ represents the expected escape one-step time for the particular site. For each site $i$, $\tau_i$ accounts for all possible transitions to neighboring states. Migration barriers $E^{\text{bond}}_{\text{mig}, ij}$, and attempt frequencies $\nu$ are incorporated to estimate escape times from each transient state:
\begin{equation}
\tau_i = \left(\nu \sum_{j=1}^{z}{\exp\left(-\frac{E^{\text{bond}}_{\text{mig}, ij}}{k_B T}\right)} \right)^{-1},
\label{eq:tau}
\end{equation}
where $\nu$ is set at $10^{13}$ Hz, as a typical atomic vibration frequency value in a crystal.

% \subsection*{Validation with Kinetic Monte Carlo simulation for the absorbing Markov chain model}
\begin{figure}[!ht]
    \centering
    \includegraphics[width=0.8\linewidth]{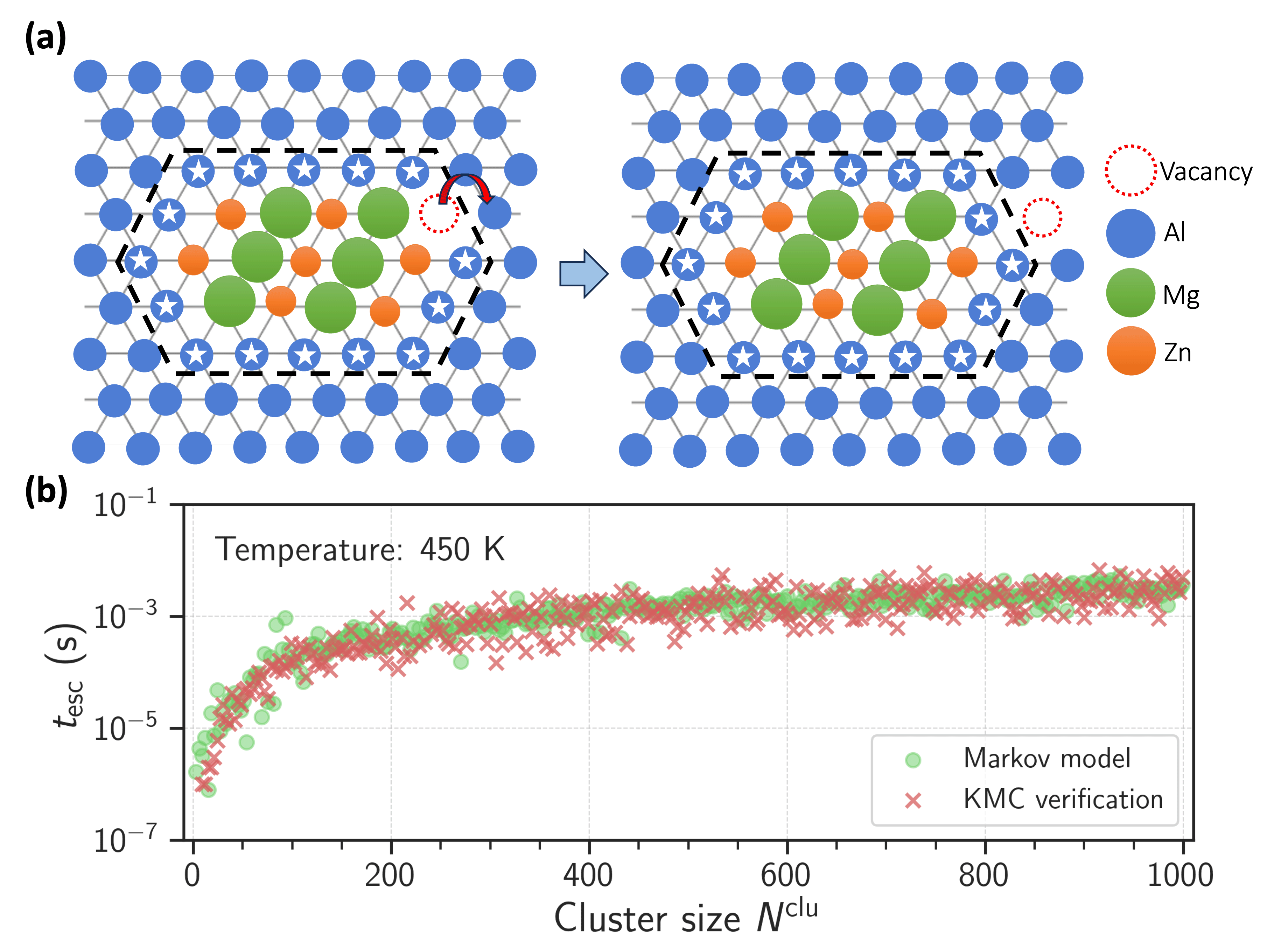}
    \caption{KMC simulations for the verification of the absorbing Markov chain model described by Eq.~\ref{eq:exit_time}. (a) Illustration of the KMC strategy used to determine the vacancy escape event from a solute cluster. Red open circles represent vacancies, while blue, green, and orange circles indicate Al, Mg, and Zn atoms, respectively. White stars mark the first nearest-neighbor shell surrounding the cluster, which is used to define the escape condition. (b) Comparison of vacancy escape times, $t_{\text{esc}}$, obtained from the analytical solution of Eq.~\ref{eq:exit_time} and from direct \ac{KMC} simulations for solute clusters generated by SA simulations with varying cluster sizes, $N^{\text{clu}}$.}
    \label{fig:kmc_verification}
\end{figure}

For solute clusters in multicomponent systems, such as Mg-Zn-rich clusters in Al-Mg-Zn alloys, we recognize that local atomic rearrangements by vacancy migration may change the actual \ac{PEL}. Despite these structural changes, we make a simplified-\ac{PEL} assumption. Specifically, we assume that the \ac{PEL} experienced by a migrating vacancy can be approximated by the static ground-state structures of the solute cluster (obtained by \ac{SA} simulations).  This simplification is reasonable for solute clusters with lower vacancy migration barriers within the clusters (as shown in Fig. \ref{fig:cluster_energies}) at lower temperatures, where low-energy cluster structures are favorable by considering both thermodynamic and kinetic factors. By making this assumption, we effectively reduce the complexity of the \ac{PEL}, enabling a computationally efficient absorbing Markov chain model to describe vacancy dynamics inside the clusters. This simplification maintains analytical tractability and computational efficiency, effectively capturing essential vacancy behavior without explicitly modeling structural rearrangements. 

We applied \ac{KMC} simulations to validate the absorbing Markov chain model and test the assumption of a simplified \ac{PEL}. \ac{KMC} and Markov chains share the same memorylessness property, which assumes that the next state depends only on the current state and not on past states. \ac{KMC} calculates transition rates and selects events probabilistically, updating the system state and time step by step. Unlike the Markov model, \ac{KMC} inherently accounts for structural rearrangements and complex \acp{PEL} by updating the lattice occupations after each vacancy migration event. 

Fig.~\ref{fig:kmc_verification} presents the verification of our absorbing Markov chain model based on the simplified \ac{PEL} of vacancy migrations inside solute clusters, which are generated and optimized from the \ac{SA} simulations (Supplementary Note 3 describes the same verification processes applied to solute clusters directly extracted from \ac{KMC} simulations). Each cluster contains one vacancy and has a fixed Zn/Mg ratio of 2 due to the \ac{SA} simulation optimization. Then, we performed \ac{KMC} simulations on these clusters, focusing on determining the real escape times required for a vacancy to leave the cluster. In \ac{KMC} simulations, the vacancies are tracked as they migrate through the cluster until they reach a site outside of the first-nearest-neighbor shell, where they are considered to have escaped from this cluster. This stopping criterion, illustrated in Fig.~\ref{fig:kmc_verification}a, mirrors the definition used in the absorbing Markov chain model, providing a consistent basis for comparison between the two approaches. The \ac{KMC} simulations were conducted at an elevated temperature of 450 K. This higher temperature facilitates fast transitions, enabling vacancies to overcome energy barriers and escape the energy basin more efficiently compared to simulations performed at room temperature. This higher temperature ensures that statistically meaningful results can be obtained within a feasible computational time. Fig.~\ref{fig:kmc_verification}b compares the escape times predicted by the absorbing Markov chain model (Eq.~\ref{eq:exit_time}) with those obtained from the \ac{KMC} simulations for clusters with varying cluster sizes, $N^{\text{clu}}$, which is the number of atoms that belong to this cluster. Here, results from absorbing Markov chain and \ac{KMC} validation exhibit strong agreement, demonstrating that the simplified \ac{PEL} approximation employed in the absorbing Markov model effectively captures the essential features of vacancy dynamics. 

\subsection*{Vacancy trapping capacity of solute clusters}

\begin{figure}[!htbp]
    \centering
    \includegraphics[width=0.8\linewidth]{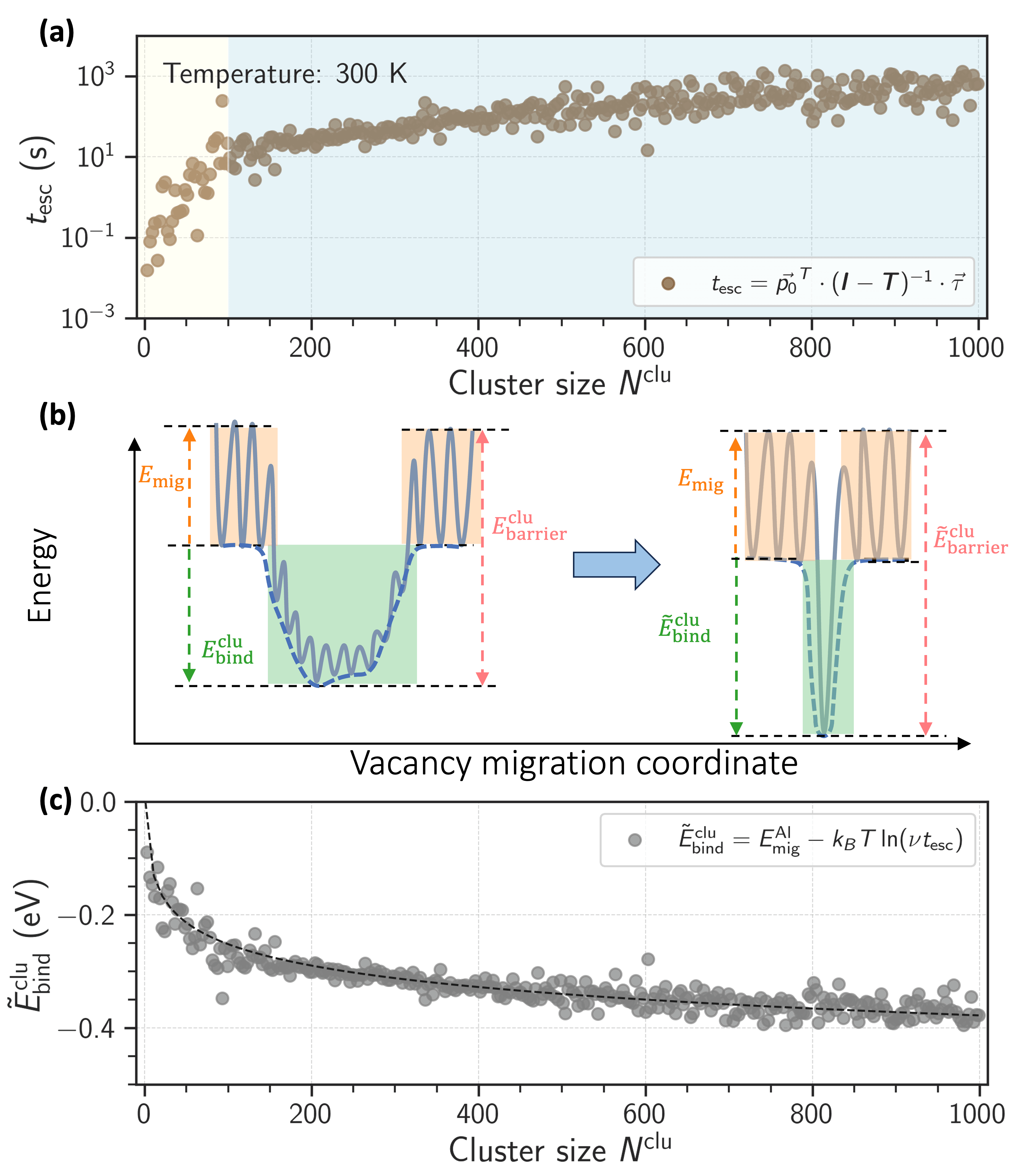}
    \caption{Analysis for vacancy trapping ability of solute clusters using the Markov chain model. (a) Vacancy escape time $t_{\text{esc}}$ calculated by Eq.~\ref{eq:exit_time} at 300~K of solute clusters in terms of cluster size $N^{\text{clu}}$. Multiple cluster configurations with the same $N^{\text{clu}}$ value were analyzed to statistically represent the trapping behavior. (b) A conceptual figure to explain the single-energy-basin approximation for a solute cluster. (c) Effective vacancy binding energy of a cluster $\tilde E^{\text{clu}}_{\text{bind}}$ defined in Eq.~\ref{eq:effective_binding_energy} in terms of cluster size $N^{\text{clu}}$. The dashed line represents a numerical fit.}
    \label{fig:trapping_ability}
\end{figure}

To investigate the influence of vacancy-cluster interactions on long-time solute clustering kinetics during natural aging, we calculated vacancy escape times, $t_{\text{esc}}$, at 300~K using Eq.~\ref{eq:exit_time} for solute clusters generated by the \ac{SA} simulations with varying $N^{\text{clu}}$. The results, shown in Fig.~\ref{fig:trapping_ability}a, reveal two distinct regimes in the dependence of $t_{\text{esc}}$ on $N^{\text{clu}}$, each reflecting different vacancy trapping behaviors. In the initial regime, as the cluster size increases, the vacancy escape time rises sharply on a logarithmic scale, from approximately $10^{-3}$~s to $10^{1}$~s. This rapid growth indicates that small clusters strongly enhance their vacancy-trapping capability as they grow. This trend is consistent with previous studies showing that the depth of the energy basin, associated with the lowest site binding energy within the cluster, increases approximately linearly with cluster size for small clusters~\cite{zurob2009model}. 

However, once the cluster size exceeds approximately 100 atoms, the trend transitions to a second regime, where further increases in $N^{\text{clu}}$ yield only modest changes in $t_{\text{esc}}$. In this regime, the vacancy trapping becomes relatively insensitive to cluster size. This behavior arises because the depth of the \ac{PEL} basin saturates, approaching the vacancy formation energy difference between the Al matrix and the solute-rich precipitate phase~\cite{vincent2006solute,soisson2007cuprecipitation,jain2023natural}. Beyond this point, additional cluster growth primarily increases the basin width rather than its depth. Although this widening effect, associated with the increased number of random walk steps a vacancy must take to escape the cluster, is generally weaker than the depth effect due to relatively low internal migration barriers, it is still non-negligible. As a result, vacancy escape times continue to increase gradually with growing cluster size.

Now, our absorbing Markov chain model captures both the depth and the width of the energy basin, incorporating the combined effects of both thermodynamic binding energies and kinetic migration barriers within the cluster. With the vacancy escape times (Fig.~\ref{fig:trapping_ability}a), we can more comprehensively characterize the vacancy trapping mechanisms in terms of energetics. We adopt a single-energy-basin approximation of the \ac{PEL}, illustrated in Fig.~\ref{fig:trapping_ability}b, where the cluster is simplified as a single site with strong vacancy binding instead of representing the cluster as a complex energy basin of the \ac{PEL}. This simplification allows us to quantify the trapping ability of the cluster through a single effective binding energy value. Thus, we define the effective cluster energy barrier, $\tilde E^{\text{clu}}_{\text{barrier}}$, which governs the escape time via the Arrhenius relationship:
\begin{equation}
\label{eq:old_escape_time}
t_{\text{esc}}=\nu^{-1}\exp\left(\frac{\tilde E^{\text{clu}}_{\text{barrier}}}{k_B T}\right).
\end{equation}
By inverting Eq.~\ref{eq:old_escape_time} and using $E_\text{mig}=E_\text{mig}^{\text{Al}}=0.58$ eV as a typical vacancy migration barrier outside of the basin, we define the effective vacancy binding energy, $\tilde E^{\text{clu}}_{\text{bind}}$, as:
\begin{equation}
\label{eq:effective_binding_energy}
\tilde E^{\text{clu}}_{\text{bind}} = E_{\text{mig}} - \tilde E^{\text{clu}}_{\text{barrier}} = E_\text{mig}^{\text{Al}} -k_B T \ln(\nu t_{\text{esc}}).
\end{equation}
As indicated by Fig.~\ref{fig:trapping_ability}b, $\tilde E^{\text{clu}}_{\text{barrier}}$ can be larger than $E_{\text{mig}}$ so that a negative value of $\tilde E^{\text{clu}}_{\text{bind}}$ indicate that a vacancy is effectively trapped by a solute cluster. Fig.~\ref{fig:trapping_ability}c plots $\tilde{E}^{\text{clu}}_{\text{bind}}$ as a function of the cluster size $N^\text{clu}$. Due to the stochastic nature of the Monte Carlo-based \ac{SA} methods and the inherent complexity of the cluster \acp{PEL}, it is challenging for individual simulations to consistently identify the lowest-energy configurations. Consequently, multiple configurations were analyzed to statistically represent the trapping behavior. A numerical fit (the dashed line in Fig.~\ref{fig:trapping_ability}b) is applied to capture the representative relationship between effective binding energies $\tilde{E}^{\text{clu}}_{\text{bind}}$ and cluster sizes, accounting for variability among different cluster configurations. These results quantitatively confirm that for large clusters, the $\tilde{E}^{\text{clu}}_{\text{bind}}$ increases rapidly and approaches a size-independent limit due to the saturation of binding strength within the interior. Nevertheless, even in this saturation stage, the effective binding energy continues to capture the impact of the random walk behavior of the vacancies within the cluster.

\subsection*{Impacts on residual vacancy concentrations for diffusion}
In multicomponent alloys, the long-time solute clustering kinetics during quenching and natural aging are strongly governed by the concentration of residual vacancies in the matrix that remain available for solute diffusion. These residual vacancy concentrations are significantly influenced by interactions between vacancies and solute atoms (or solute clusters), which can effectively trap vacancies and thus reduce the number of mobile vacancies in the matrix. The effective vacancy binding energy to a given solute cluster, denoted as $\tilde{E}^{\text{clu}}_{\text{bind}}$ in Eq.~\ref{eq:effective_binding_energy}, provides a rigorous and quantitative framework to evaluate the influence of various solute cluster types and sizes on residual vacancy concentrations as described in the following analyses.

Based on previous studies~\cite{vincent2006solute,vincent2008precipitation,soisson2007cuprecipitation,yang2021natural,francis2016microalloying,wang2017precipitation}, we classify the equilibrium vacancy concentration as a function of temperature, $c^{\text{vac}}_{\text{eq}}\left(T\right)$, into three distinct categories, which are described as follows:

\begin{equation}
\label{eq:vac_eq_total}
c^{\text{vac}}_{\text{eq}}\left(T\right) =  c^{\text{vac,X}}_{\text{eq}}  + c^{\text{vac,clu}}_{\text{eq}} + c^{\text{vac,res}}_{\text{eq}}.
\end{equation}

The first group, $c^{\text{vac,X}}_{\text{eq}}$, denotes the atomic fraction of vacancies trapped by individual solute atom X isolated in the Al matrix as follows: 
\begin{equation}
c^{\text{vac,X}}_{\text{eq}}\left(T\right) = c^{\text{vac,Al}}_{\text{eq}} \left(T\right)\sum_\text{X}\left[z c^{\text{X}}_{1}  \exp\left(-\frac{E_{\text{bind}}^{{\text{X}}}}{k_BT} \right) \right],
\end{equation}
The reference vacancy concentration in pure Al is represented by $c^{\text{vac,Al}}_{\text{eq}}$. Both this quantity and the concentration of isolated solute atoms, $c^{\text{X}}_{1}$, are expressed as atomic fractions. The parameter $z$ represents the coordination number, i.e., the number of nearest-neighbor sites surrounding a solute atom. The term $E_{\text{bind}}^{\text{X}}$ denotes the vacancy binding energy associated with this individual solute atom X, where a negative value signifies attractive binding.

The second group, $c^{\text{vac,clu}}_{\text{eq}}$, is the fraction of vacancies trapped by solute clusters as follows: 

\begin{equation}
c^{\text{vac,clu}}_{\text{eq}}\left(T\right) = c^{\text{vac,Al}}_{\text{eq}}\left(T\right) \sum_\text{clu}\left[\left(z^{\text{clu}}+N^{\text{clu}}\right) c^{\text{clu}}  \exp\left(-\frac{\tilde E_{\text{bind}}^{\text{clu}}}{k_BT}\right) \right], 
\end{equation}
The sizes of these clusters are expressed as $N^{\text{clu}}$, which is the number of lattice sites occupied by this cluster. The concentrations of these clusters, $c^{\text{clu}}$, are expressed as an atomic fraction (the number of clusters in the whole simulation supercell divided by the number of lattice sites in this supercell). The term $z^{\text{clu}}$ represents the number of neighboring sites that are direct nearest neighbors to any atom belonging to the cluster. These sites provide additional sites for trapping vacancies at the cluster-matrix interfaces. $\tilde E_{\text{bind}}^{\text{clu}}$ is the effective vacancy binding energy of clusters calculated by Eq.~\ref{eq:effective_binding_energy}, with negative values indicating attractive binding.

The last group, $c^{\text{vac,res}}_{\text{eq}}$, represents the residual vacancies that remain freely distributed in the Al matrix.
\begin{equation}
\label{eq:vac_eq_res}
c^{\text{vac,res}}_{\text{eq}}\left(T\right) =  c^{\text{vac,Al}}_{\text{eq}}\left(T\right) \left[1-\sum_\text{X}{\left(z+1\right) c^{\text{X}}_{1}} - \sum_\text{clu}\left(z^{\text{clu}}+N^{\text{clu}}\right) c^{\text{clu}} \right],
\end{equation}

Although Eq.~\ref{eq:vac_eq_res} offers a refined estimate of the equilibrium residual vacancy concentration in the Al matrix, the vacancy concentration during quenching remains far from equilibrium due to rapid kinetic processes. Recent computational and experimental studies~\cite{wang2024modelling,chen2025using} have demonstrated that vacancy supersaturation can persist for several hours to days ($\sim10^{5}$-$10^{6}$ seconds) in regions far from \acp{GB}. Therefore, we focus on the interactions between vacancies and both solute atoms and solute clusters, while assuming that the total vacancy concentration remains constant during quenching in large-grained regions. The expression for the dynamic vacancy concentration is given by:
\begin{equation}
\label{eq:vac_dyn_res1}
c^{\text{vac,res}}_{\text{dyn}} \left(T\right)= c^{\text{vac}}_{\text{eq}}\left(800K\right) \frac{\sum_\text{Al} \exp\left({-\frac{E_{\text{bind}}^{\text{site}}}{k_B T}}\right)}{\sum_\text{all} \exp\left({-\frac{E_{\text{bind}}^{\text{site}}}{k_B T}}\right)},
\end{equation}
where the numerator and denominator represent the total probability of finding vacancies in all lattice sites of the Al matrix and in all lattice sites of the entire alloy, respectively, with $E_{\text{bind}}^{\text{site}}$ as the vacancy binding energy at the corresponding site. It is important to note that the local binding energy of sites in the pure Al matrix is defined to be zero. By substituting the effective cluster binding energy, $\tilde{E}_{\text{bind}}^{\text{clu}}$, for $E_{\text{bind}}^{\text{site}}$ for each site belong to clusters, Eq.~\ref{eq:vac_dyn_res1} can be reformulated as:

\begin{equation}
\label{eq:vac_dyn_res2}
c^{\text{vac,res}}_{\text{dyn}} \left(T\right)= 
 \frac{ c^{\text{vac}}_{\text{eq}}\left(800K\right)\left(1-\sum_\text{X}{\left(z+1\right) c^{\text{X}}_{1}} - \sum_\text{clu}\left(z^{\text{clu}}+N^{\text{clu}}\right) c^{\text{clu}}\right)}{1-\sum_\text{X}{\left[\left(z+1\right)c^{\text{X}}_{1}-zc^{\text{X}}_{1}\exp\left(-\frac{E_{\text{bind}}^{{\text{X}}}}{k_BT} \right)\right]} - \sum_\text{clu}{\left[\left(z^{\text{clu}}+N^{\text{clu}}\right)c^{\text{clu}}\left(1-\exp\left(-\frac{\tilde E_{\text{bind}}^{\text{clu}}}{k_BT}\right)\right)\right]}}
 .
\end{equation}
The term $c^{\text{vac,res}}_{\text{dyn}}$ represents the concentration of mobile vacancies remaining in the Al matrix that are available to facilitate solute diffusion. This model enables more accurate prediction of vacancy concentrations, particularly in systems containing large solute clusters.

\begin{figure}[!ht]
    \centering
    \includegraphics[width=0.8\linewidth]{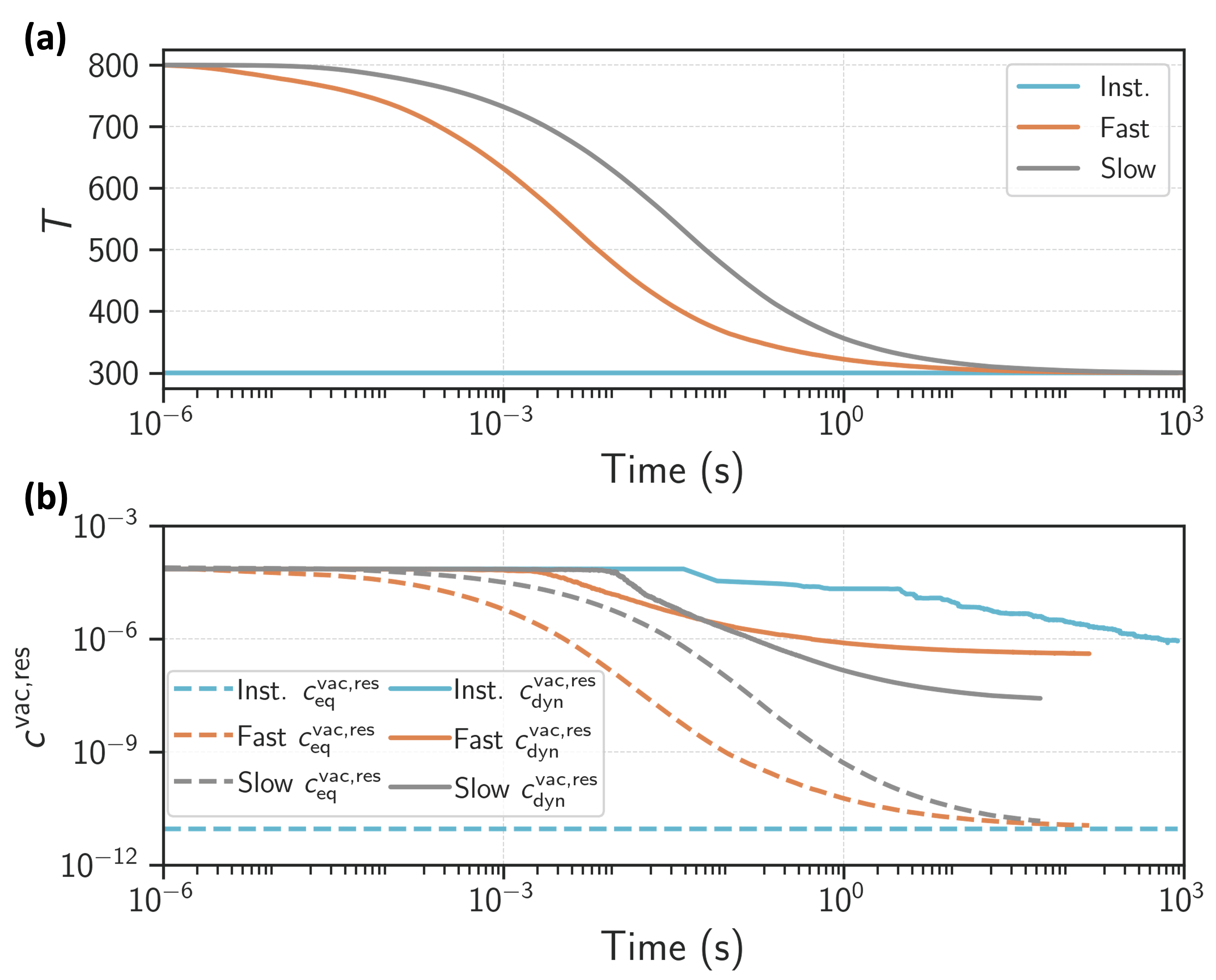}
    \caption{Evolution of residual vacancy concentration during quenching under three different cooling rates, where the corresponding cluster structures and effective vacancy-cluster interaction energies estimated from KMC simulations and the Markov chain model. (a) Temperature profiles for the three cooling conditions used in the KMC simulations. (b) Time evolution of the equilibrium residual vacancy concentration ($c^{\text{vac,res}}_{\text{eq}}$, defined in Eq.~\ref{eq:vac_eq_res}, shown as dashed lines) and the dynamic residual vacancy concentration ($c^{\text{vac,res}}_{\text{dyn}}$, defined in Eq.~\ref{eq:vac_dyn_res2}, shown as solid lines) for the instantaneous, fast, and slow cooling scenarios presented in (a).}
    \label{fig:kmc_cv}
\end{figure}

Fig.~\ref{fig:kmc_cv}a displays the temperature evolution profiles under three different cooling conditions: instantaneous quenching, fast cooling, and slow cooling (see Supplementary Note 4 for cluster snapshots). Here, cluster structures and effective vacancy-cluster interaction energies were calculated from KMC simulations and the Markov chain model described above. Fig.~\ref{fig:kmc_cv}b presents the corresponding evolution of residual vacancy concentrations in the Al matrix, including both the equilibrium concentration $c^{\text{vac,res}}_{\text{eq}}$ (from Eq.~\ref{eq:vac_eq_res}) and the dynamic concentration $c^{\text{vac,res}}_{\text{dyn}}$ (from Eq.~\ref{eq:vac_dyn_res2}) under these three conditions. In all cases, $c^{\text{vac,res}}_{\text{dyn}}$ remains higher than $c^{\text{vac,res}}_{\text{eq}}$ because equilibrium has not yet been reached. For the instantaneous quenching case, solute clusters cannot grow rapidly, leading to fewer large clusters capable of effectively trapping vacancies. As a result, a higher $c^{\text{vac,res}}_{\text{dyn}}$ is observed in the Al matrix. Over time, as clustering progresses and larger clusters form, $c^{\text{vac,res}}_{\text{dyn}}$ gradually decreases. In contrast, under more realistic cooling conditions (fast and slow cooling), stronger vacancy-cluster interactions are observed due to the formation of larger clusters during the cooling process. These clusters absorb more vacancies through strong binding interactions, resulting in lower $c^{\text{vac,res}}_{\text{dyn}}$ at a given time. Overall, $c^{\text{vac,res}}_{\text{dyn}}$ tends to be higher for faster cooling rates than for slower ones, as faster cooling suppresses the formation of additional large clusters that trap vacancies more effectively. Interestingly, this trend is opposite to that observed for $c^{\text{vac,res}}_{\text{eq}}$, where faster cooling corresponds to a lower temperature at a given time, thereby yielding lower equilibrium vacancy concentrations. This opposite trend can be rationalized by noting that faster cooling rates lead to a larger deviation between the dynamic and equilibrium vacancy concentrations, i.e., a greater gap between $c^{\text{vac,res}}_{\text{dyn}}$ and $c^{\text{vac,clu}}_{\text{eq}}$, due to the system being driven further from equilibrium by rapid kinetics.

\subsection*{Cluster dynamics simulations for long-time clustering kinetics}

Predictions of $c^{\text{vac,res}}_{\text{dyn}}$, the concentration of mobile vacancies available for solute diffusion in the Al matrix, provide a solid foundation for studying the long-time solute clustering kinetics in Al alloys. The diffusivity of solute species, $D^{\text{X}}$, is directly influenced by vacancy concentration and migration barriers, and follows the relationship $D^{\text{X}} \propto c^{\text{vac,res}}_{\text{dyn}} \exp \left(-{E_{\text{mig}}^{\text{X}}}/{k_B T}\right)$. This diffusivity governs both the nucleation and growth rates, which are critical for describing precipitation kinetics. The \ac{KMC} results shown in Fig.~\ref{fig:trapping_in_kmc} demonstrate that traditional \ac{KMC} simulations encounter challenges in efficiently simulating cluster growth beyond the early stages. These limitations arise due to strong vacancy-cluster interactions and the finite number of vacancies in a typical simulation supercell. To overcome these constraints and to study clustering behavior over longer timescales during natural aging, \ac{CD} simulations were employed. By incorporating the accurate vacancy supersaturation obtained from the absorbing Markov chain model, the \ac{CD} framework (see Methods section for details) bridges atomistic-scale vacancy-cluster interaction energetics with macroscopic precipitation processes, enabling robust modeling of cluster size distribution evolution over extended time periods.

\begin{figure}[!ht]
    \centering
    \includegraphics[width=0.8\linewidth]{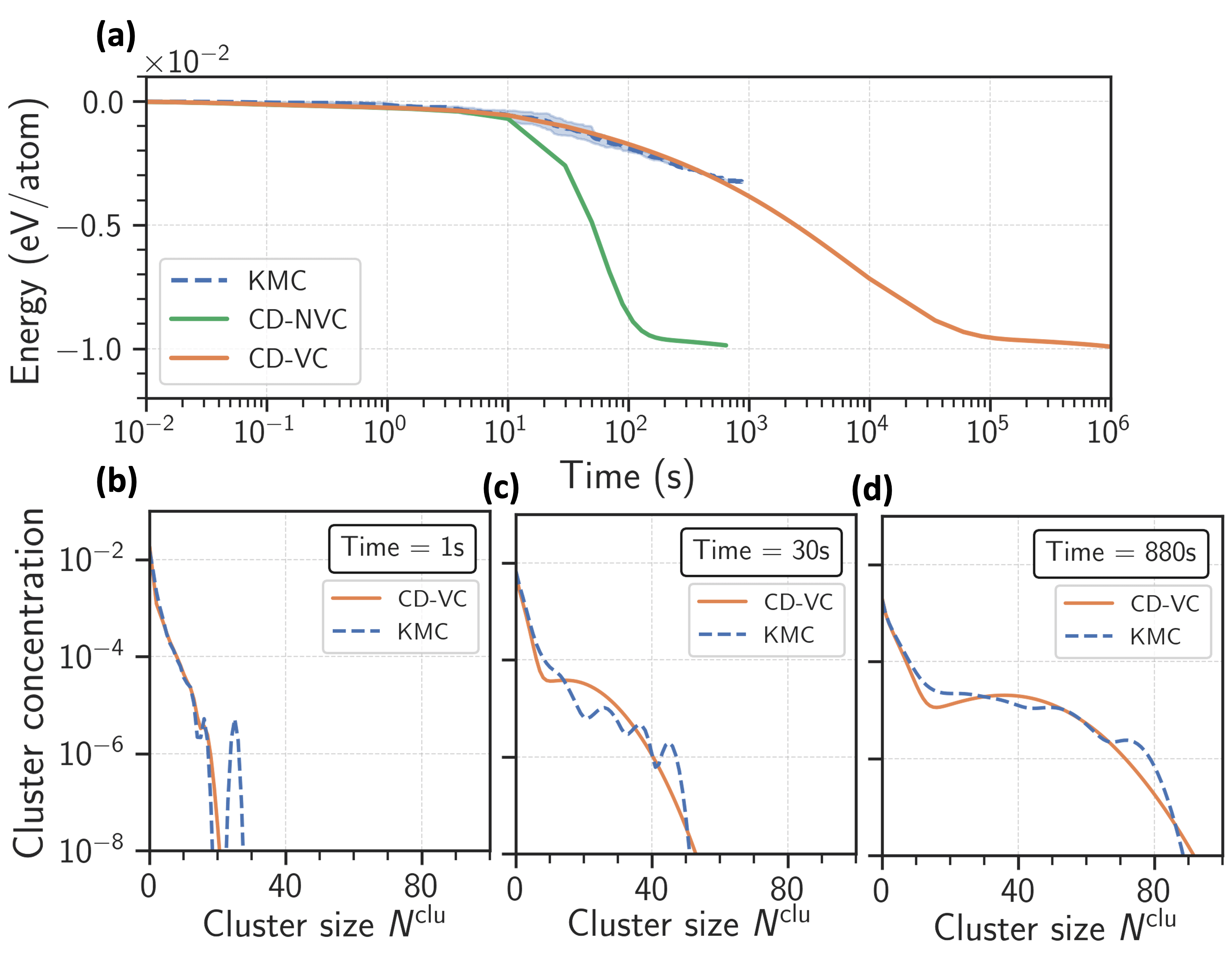}
    \caption{Comparison between CD and KMC simulations under infinitely fast (instantaneous) cooling conditions. (a) Time evolution of system energy per atom. Results are shown for KMC (blue dashed line with light blue shades that represent the standard errors from three individual KMC simulations), CD without vacancy-cluster interaction corrections (CD-NVC, green solid line), and CD with vacancy-cluster interaction corrections (CD-VC, orange solid line). (b)-(d) The distributions of cluster concentrations $c^{\text{clu}}$ as a function of cluster size $N^{\text{clu}}$ at the time of (b) 1 s, (c) 30 s, and (d) 880 s after the starting of the simulations predicted by both KMC (blue dashed lines) and CD-VC (orange solid lines) method.}
    \label{fig:cd_kmc}
\end{figure}

To validate the predictive capability of the \ac{CD} model, particularly its incorporation of vacancy-cluster interaction corrections in solute diffusivity, we directly compare the \ac{CD} results with \ac{KMC} simulations under infinitely fast (instantaneous) cooling conditions. Fig.~\ref{fig:cd_kmc}a presents the time evolution of the system energy per atom, showing three datasets: \ac{KMC} simulations, \ac{CD} without vacancy-cluster interaction corrections (CD-NVC), and \ac{CD} with corrections (CD-VC). The CD-VC model includes the effects of the dynamic residual vacancy concentration, $c^{\text{vac,res}}_{\text{dyn}}$, which captures vacancy supersaturation and its influence on solute diffusivity $D^{\text{X}}$. Since solute diffusivity directly governs the condensation and evaporation rates in the \ac{CD} framework, accurately incorporating $c^{\text{vac,res}}_{\text{dyn}}$ allows CD-VC simulations to closely reproduce the energy evolution obtained from \ac{KMC} over time. In addition, Figs.~\ref{fig:cd_kmc}b-d compare the distributions of cluster concentrations, $c^{\text{clu}}$, as a function of cluster size, $N^{\text{clu}}$, obtained from CD-VC predictions and \ac{KMC} simulations at different time points: 1~s, 30~s, and 880~s after the starting of the simulations. These comparisons demonstrate strong agreement between the CD-VC and \ac{KMC} models in predicting the temporal evolution of cluster size distributions across a wide range of times and temperatures.

\begin{figure}[!ht]
    \centering
    \includegraphics[width=0.7\linewidth]{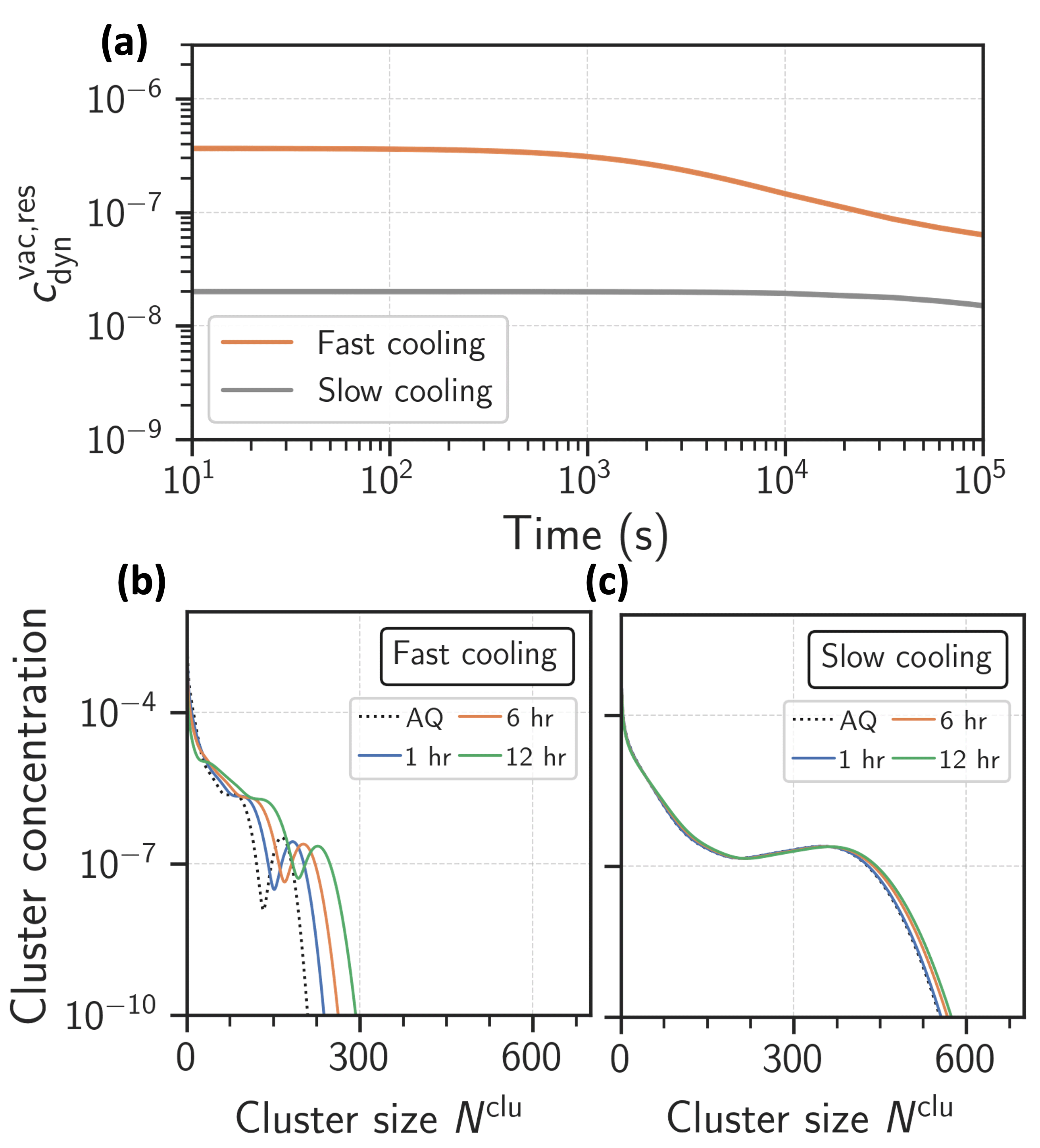}
    \caption{Long-time CD simulation results of natural aging solute clustering for alloys obtained from different cooling conditions. (a) Residual vacancy concentration evolution during natural aging from different cooling conditions. (b)-(c) The distributions of cluster concentrations $c^{\text{clu}}$ as a function of cluster size $N^{\text{clu}}$ for various natural aging time (as-quenched(10 seconds),  1 hour, 6 hours, and 12 hours) for the alloys obtained from fast cooling (b) and slow cooling (c) conditions. The temperature profiles for the fast and slow cooling cases are described in Fig.~\ref{fig:kmc_cv}(a).}
    \label{fig:cd_long}
\end{figure}

Finally, the combination of \ac{KMC} simulations and the \ac{CD} model, informed by the calculated dynamic vacancy concentrations $c^{\text{vac,res}}_{\text{dyn}}$, is employed to simulate the long-time natural aging process following various cooling conditions, beginning from the as-quenched state. In this hybrid approach, \ac{KMC} simulations are first used to model the quenching process until the temperature reaches approximately 300~K (corresponding to 10 seconds after the starting of the simulations in our cases). At this point, the system energy evolution plateaus as shown in Fig.~\ref{fig:trapping_in_kmc}, indicating that vacancies in the supercell are effectively trapped by solute clusters, and the system reaches a metastable state. Subsequently, the \ac{CD} model is applied to simulate the long-term behavior of solute cluster evolution during natural aging. 

Fig.~\ref{fig:cd_long}a presents the evolution of residual vacancy concentrations in the Al matrix, $c^{\text{vac,res}}_{\text{dyn}}$, during natural aging at 300~K for both fast and slow cooling scenarios. These scenarios correspond to the two temperature profiles shown in Fig.~\ref{fig:kmc_cv}a. The initial value of $c^{\text{vac,res}}_{\text{dyn}}$ at the onset of aging (i.e., at time $t = 10$~s, marking the end of quenching) strongly depends on the cooling rate. Faster cooling limits the time available for solute clusters to form and trap vacancies, resulting in higher initial values of $c^{\text{vac,res}}_{\text{dyn}}$, which are consistent with the values of $c^{\text{vac,res}}_{\text{dyn}}$ shown in Fig.~\ref{fig:kmc_cv}b for these two cooling cases.

During the subsequent natural aging process, $c^{\text{vac,res}}_{\text{dyn}}$ continues to decrease over time for both cooling scenarios. This decline reflects the gradual annihilation of excess vacancies at various sinks, particularly growing solute clusters. Although dislocations and \acp{GB} are also known to act as vacancy sinks, they typically become significant only at much longer timescales (on the order of $10^4$-$10^5$~s after quenching~\cite{wang2024modelling}) and are therefore not explicitly included in the present model. The time-dependent evolution of $c^{\text{vac,res}}_{\text{dyn}}$ plays a critical role in determining solute diffusivity and thus strongly influences the kinetics of solute clustering during natural aging. Figs~\ref{fig:cd_long}b and \ref{fig:cd_long}c show the cluster concentration distributions, $c^{\text{clu}}$, as a function of cluster size, $N^{\text{clu}}$, predicted by the \ac{CD} simulations at various natural aging times: as-quenched (AQ), 1 hour, 6 hours, and 12 hours after the quenching. These results are shown for both fast and slow cooling conditions as shown in Fig.~\ref{fig:kmc_cv}a. Under the fast cooling condition (Fig.~\ref{fig:cd_long}b), $c^{\text{clu}}(N^{\text{clu}})$ distributions corresponding to small quenched-in clusters are initially present. These clusters continue to grow and coarsen during natural aging due to the elevated diffusivity enabled by higher vacancy supersaturation in the Al matrix. In contrast, the slow cooling condition (Fig.~\ref{fig:cd_long}c) leads to the formation of larger initial as-quenched clusters and a substantially reduced concentration of mobile vacancies. As a result, the cluster size distribution evolves much more slowly during natural aging. These results demonstrate that the rate of cluster evolution during aging is strongly influenced by the preceding thermal history, with faster cooling leading to more rapid diffusion and clustering kinetics due to a greater degree of vacancy supersaturation.

\section*{Discussion}

The absorbing Markov chain model developed in this work represents a significant advancement in understanding vacancy-cluster interactions, particularly in reconciling previously conflicting descriptions of effective vacancy binding energies. As mentioned in the Introduction, earlier studies~\cite{zurob2009model,clouet2004nucleation,sha2005kinetic,liang2012kinetics,soisson1996monte,soisson2000monte,soisson2007cuprecipitation,soisson2010atomistic,soisson2022atomistic,vincent2006solute,vincent2008precipitation,wang2017precipitation,jain2023natural} have provided contrasting models for how the binding energy evolves with cluster size. For instance, Zurob and Seyedrezai~\cite{zurob2009model} proposed a linear relationship between vacancy binding energy and cluster size, suggesting $E^{\text{clu}}_{\text{bind}} \propto N^{\text{clu}} E^{\text{X}}_{\text{bind}}$, where $N^{\text{clu}}$ is the number of solute atoms in the cluster and $E^{\text{X}}_{\text{bind}}$ is the vacancy binding energy of an individual solute atom. In contrast, Soisson \textit{et al.}~\cite{soisson2007cuprecipitation} and Vincent \textit{et al.}~\cite{vincent2008precipitation} studied Cu clusters in $\alpha$-Fe and found that the effective binding energy saturates to a value determined by the difference in vacancy formation energies between the matrix and the precipitate: $E^{\text{clu}}_{\text{bind}} = E_{\text{form}}^{\text{vac,ppt}} - E_{\text{form}}^{\text{vac,mx}}$.

Our results, derived from the absorbing Markov chain model (Fig.~\ref{fig:trapping_ability}), reveal a continuous size-dependent transition in the effective vacancy binding energy, $\tilde{E}^{\text{clu}}_{\text{bind}}$, that naturally reconciles these perspectives. For small clusters (up to roughly 100 atoms), $\tilde{E}^{\text{clu}}_{\text{bind}}$ increases sharply with cluster size, reflecting a rapid deepening of the ``vacancy prison'' energy basin as more solute atoms contribute binding sites. This regime aligns with the linear behavior described by Zurob and Seyedrezai~\cite{zurob2009model}. As the cluster size increases further, the growth of $\tilde{E}^{\text{clu}}_{\text{bind}}$ slows and gradually saturates, consistent with the mechanism proposed by Soisson \textit{et al.}~\cite{soisson2007cuprecipitation} and Vincent \textit{et al.}~\cite{vincent2008precipitation}, where the binding energy reflects the intrinsic difference in vacancy formation energies between the cluster and the matrix. Importantly, our model also captures the continued, albeit slower, increase in the vacancy escape time, $t_{\text{esc}}$, and thus in $\tilde{E}^{\text{clu}}_{\text{bind}}$, even in the saturation regime. This increase arises from the growing number of random walk steps required for a vacancy to diffuse out from the core of larger clusters. Overall, this continuous transition, inherently captured by the absorbing Markov chain model, demonstrates that previous models effectively described different stages of the same underlying physical phenomenon.

The Markov chain model developed in this study plays a crucial role in enabling mesoscale simulations of early-stage precipitation kinetics. A key challenge in accurately modeling this behavior lies in quantifying the dynamic concentration of vacancies in the presence of solute clusters formed during quenching. This is addressed by incorporating the effective vacancy binding energies (Eq.~\ref{eq:vac_dyn_res2}) into the \ac{CD} framework (Fig.~\ref{fig:kmc_cv}). Our predicted vacancy concentrations show good agreement with experimental observations, which consistently report that faster cooling rates lead to higher residual vacancy concentrations~\cite{liu2015early,yang2021natural}. This vacancy concentration correction is essential, as it directly governs solute diffusivity, nucleation rate, and growth kinetics during aging~\cite{wang2024modelling,chen2025using,myhr2024modeling}. The predictive capability of this approach is validated by excellent agreement with direct \ac{KMC} simulations (Fig.~\ref{fig:cd_kmc}). Furthermore, long-time \ac{CD} simulations (Fig.~\ref{fig:cd_long}) reveal that quenching conditions have a lasting impact on the rate of cluster evolution during natural aging. This result provides a quantitative explanation for extensive experimental findings that slow cooling rates degrade mechanical properties such as strength and toughness~\cite{thomas1994quench,dons1983quench,milkereit2015quench,strobel2011relating,thompson1971quench,sahu2025situ}. The degradation arises because slow cooling promotes the formation of coarse, quench-induced precipitates, which consume both solute atoms and excess vacancies, limiting the formation of fine, strengthening precipitates during subsequent aging~\cite{deschamps2009influence,liu2015early,yang2021natural}.

Despite the advances enabled by our integrated multiscale framework, some limitations remain in the current implementation. In particular, our \ac{CD} simulations, although informed by \ac{KMC} and Markov chain model results, are performed using a fixed Zn/Mg ratio of 2. This choice dose not reflect inherent limitations of \ac{CD} method, as well-developed theories for modeling clustering in non-stoichiometric compounds already exist for decades~\cite{reiss1950kinetics,hirschfelder1974kinetics,stauffer1976kinetic,trinkaus1983theory,wu1993general,clouet2009modeling}. Instead, this ratio was adopted to avoid the computationally prohibitive task of exploring the full two-dimensional compositional space defined by the number of Mg and Zn atoms ($n_\text{Mg}$, $n_\text{Zn}$) in cluster formation enthalpy calculations. The selected ratio is physically meaningful, corresponding to the minimum cluster formation enthalpy pathway (see Supplementary Note 2) and consistent with the Zn/Mg ratio observed in large \ac{GP}-I zones after prolonged aging~\cite{lervik2021atomic}, as well as with
recent thermodynamic evaluations of Mg-Zn clustering in Al alloys~\cite{liu2024revisiting}. While this approach does not capture the dynamic evolution of the Zn/Mg ratio, which is known to vary with alloy composition and thermomechanical processing conditions~\cite{thronsen2023evolution,liu2020formation,geng2022earlystage}, it remains a reasonable approximation within the scope of this study. Specifically, our focus is on vacancy trapping effects and their impact on long-term aging kinetics, where this simplification is justified by two key observations. First, at the atomistic level, we find that vacancy trapping behavior is comparable between clusters derived from \ac{SA} simulations with a fixed Zn/Mg ratio of 2 and those from \ac{KMC} simulations with varying Zn/Mg ratios (see Supplementary Note 3). Second, at the mesoscale, our simplified \ac{CD} model accurately reproduces the system energy evolution and cluster size distributions predicted by direct \ac{KMC} simulations (Fig.~\ref{fig:cd_kmc}). These results collectively support the validity of the fixed-ratio approximation for the purposes of this study.

Additionally, while our \ac{CD} simulations effectively model long-term cluster evolution kinetics, they remain inherently limited by assumptions from \ac{CNT} and simplified growth models, such as single-atom addition mechanisms. These simplifications restrict the ability of \ac{CD} to capture complex phenomena such as coarsening, cluster morphology evolution, and microstructural interactions. To overcome these limitations, larger-scale simulation techniques, such as phase field (\ac{PF}) modeling~\cite{li1998computer,vaithyanathan2002multiscale,kleiven2020precipitate}, are essential. \ac{PF} methods are well-suited for simulating microstructural evolution and morphological changes across larger spatial and temporal scales. The predictive capability of \ac{PF} simulations, however, depends critically on the accurate parameterization of the free energy functional and mobility terms. Our integrated computational framework provides key inputs for such models: surrogate models trained on first-principles data yield reliable thermodynamic quantities for free energy landscapes, while the dynamic vacancy concentrations derived from our approach offer corrections to solute diffusivities and mobilities required in \ac{PF} formulations. This integration opens pathways for more comprehensive, large-scale simulations of precipitation and microstructure evolution in engineering alloys.

In conclusion, the absorbing Markov chain framework presented here successfully bridges atomistic vacancy interaction mechanisms with macroscopic precipitation behavior. The derived insights into vacancy-cluster interactions and the refined modeling of dynamic vacancy concentrations establish a robust foundation for understanding clustering kinetics, predicting precipitation behaviors, optimizing heat treatments, and ultimately enhancing the performance of engineering alloys such as the Al-Mg-Zn system.

\section*{Methods}

\subsection*{Kinetic Monte Carlo simulations}
The details of the \ac{KMC} simulation methods are provided in our recent studies and are briefly summarized below~\cite{xi2024kinetic}. The most common approach to \ac{KMC} simulations is the first-order residence time algorithm~\cite{bortz1975new}. In this algorithm, we consider $z=12$ first nearest neighbors of a vacancy. Assuming each neighboring atom vibrates at a typical crystal frequency ($\nu_i = 1 \times 10^{13}$ Hz), it results in a transition probability per unit time $r_i = \nu_i \exp(-E^{\text{bond}}_{\text{mig}, i}/k_B T)$, with $E^{\text{bond}}_{\text{mig}, i}$ representing the migration energy barrier. However, low-barrier events near the solute clusters often lead to forward and backward vacancy jumps, causing frequent 'flicker' events that considerably slow down first-order \ac{KMC} algorithms, particularly at low temperatures. We employed a second-order residence time algorithm~\cite{athenes1997identification,athenes2000effects,mason2004stochastic}, which bypasses flicker events by considering two-step vacancy jumps. For solute clustering studies, \ac{KMC} simulations were performed in a $30 \times 30 \times 30$ \ac{FCC} supercell of Al-2.86 at.\%Mg-2.38 at.\%Zn (one vacancy in a whole simulation supercell), representative of 7075 Al alloy compositions. The solute concentrations are within the range of compositions of the 7075 Al alloys. The initial configurations were equilibrated at 800 K (typical solutionizing temperature for Al alloys) via \ac{CMC} for approximately 10$^9$ steps. During simulated quenching, temperatures were updated at each timestep according to predefined cooling profiles. For the absorbing Markov chain model verification, simulations employed the same $17 \times 17 \times 17$ \ac{FCC} supercell containing clusters obtained from \ac{SA} simulations and one vacancy, conducted at a higher temperature of 450 K for simulation efficiency.

\subsection*{Absorbing Markov chain}

\begin{figure}[!htbp]
    \centering
    \includegraphics[width=0.7\linewidth]{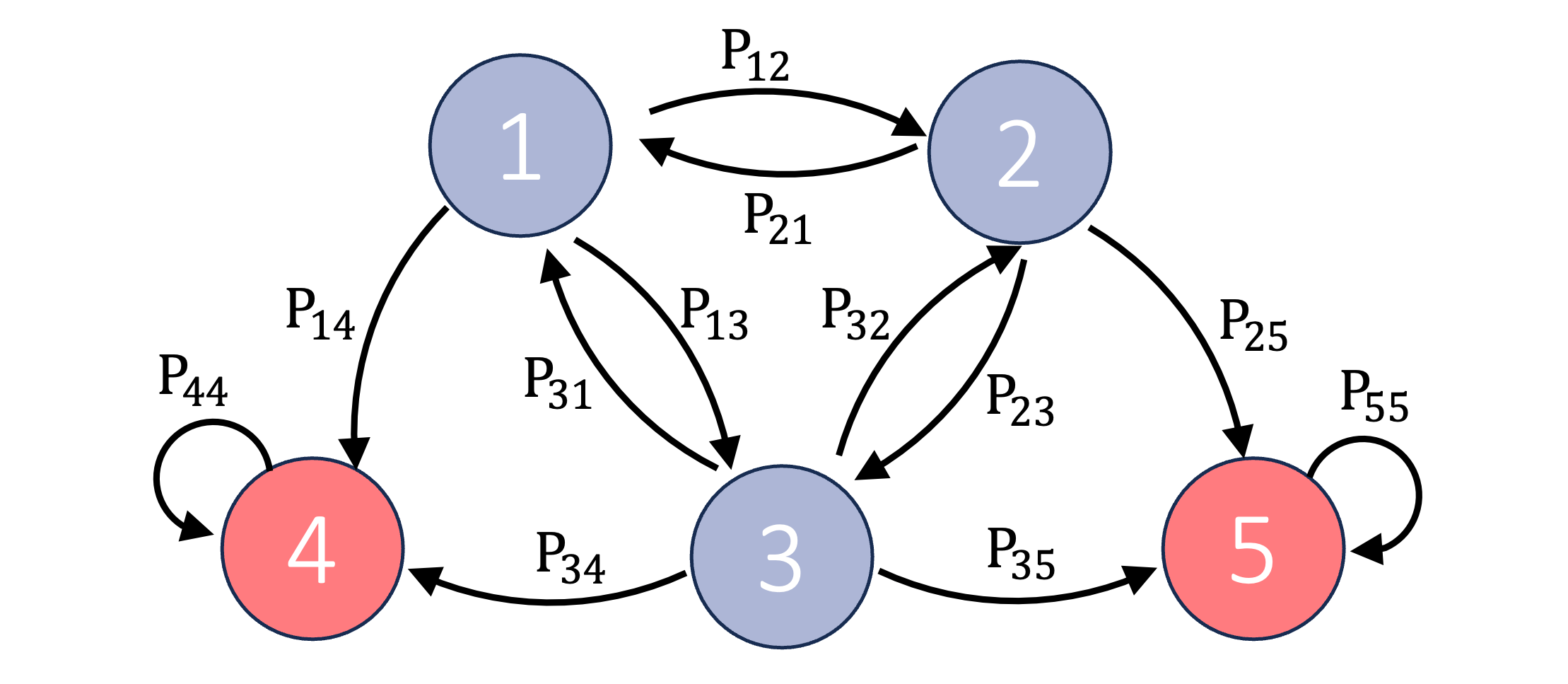}
    \caption{A simplified example of the transition graph of an absorbing Markov chain with 5 states, where states 1, 2, and 3 are transient states and states 4 and 5 are absorbing states.}
    \label{fig:markov_chain}
\end{figure}

A Markov chain is a model that describes a system transitioning between a set of states with certain probabilities. It assumes that the next state depends only on the current state, not the past sequence of states. Mathematically, it is fully defined by a set of possible states and a transition matrix, whose elements represent the probabilities of moving from a given current state to each possible next state. The transition matrix $\pmb{P}$ is a matrix of all transition rates $P_{ij}$, which is the probability for state $i$ to state $j$. The summation of all elements in each row should be 1, we have $\sum_j P_{ij} = 1$.

%\ac{MC} simulations closely relate to Markov chains as both model stochastic system evolution via states with Markovian transition probabilities. (I delete this sentence because it is not very helpful

An absorbing Markov chain is a type of Markov chain that includes at least one absorbing state, which is defined as a state that, once entered, cannot be exited. This means the probability of an absorbing state remaining in the absorbing state is 1, i.e., $P_{ii}=1$, and the probability of this absorbing state transitioning to any other state is 0, i.e., $P_{ij}=0$ for $i \neq j$. In contrast, a transient state is a state that can be exited, making it possible to leave and potentially never return. The transition matrix $\pmb P$ of an absorbing Markov chain can be expressed in canonical form as:
\begin{equation}
    \label{eq:markov_matrix}
    \pmb P= \begin{pmatrix}
\pmb  T & \pmb  R \\
\pmb  0 & \pmb  I
\end{pmatrix},
\end{equation}
where $\pmb I$ is a $n_\text{ab} \times n_\text{ab}$ identity matrix corresponding to the absorbing states, with $n_\text{ab}$ is the number of absorbing states. $\pmb T$ is a $n_\text{tr} \times n_\text{tr}$ transient matrix containing the probabilities of transitions between transient states. $n_\text{tr}$ is the number of transient states. $\pmb R$ is a $n_\text{tr} \times n_\text{ab}$ matrix, representing transition probabilities from transient to absorbing state. $\pmb 0$ is a zero $n_\text{ab} \times n_\text{tr}$ matrix indicating no transitions from absorbing to transient states.

Fig.~\ref{fig:markov_chain} shows a simplified example graph of an absorbing Markov chain with 5 states, where states 1, 2, and 3 are transition states and states 4 and 5 are absorbing states.  Eq.~\ref{eq:markov_matrix} for this specific graph is:
\begin{equation}
    \label{eq:markov_matrix_example}
    \pmb{P}_\text{example}= \begin{pmatrix}
    0 & P_{12} & P_{13} & P_{14} & 0 \\ 
    P_{21} & 0 & P_{23} & 0 & P_{25} \\ 
    P_{31} & P_{32} & 0 & P_{34} & P_{35} \\
    0 & 0 & 0 & P_{44} & 0 \\ 
    0 & 0 & 0 & 0 & P_{55}
    \end{pmatrix},
\end{equation}
where $P_{44}=P_{55}=1$ because of the absorbing states.

\subsection*{Cluster dynamics simulations}
The \ac{CD} simulations bridge atomistic simulations and macroscopic approaches by tracking the evolution of clusters in an alloy~\cite{clouet2005precipitation,clouet2009modeling,lepinoux2009modelling,lepinoux2010comparing,lepinoux2018multiscale}. In this formalism, clusters with a size of $N^{\text{clu}} = n$ atoms were explicitly tracked via their concentration $c_n$, governed by a set of coupled ordinary differential equations known as the master equations:
\begin{equation}
\frac{dc_n}{dt} = (\alpha_{n+1} c_{n+1} + \beta_{n-1} c_{n-1}) - (\alpha_n + \beta_n) c_n, \quad\text{for } n \geq 2,
\end{equation}
where $\alpha_n$ represents the evaporation rate (the rate of cluster shrinkage from size $n$ to $n - 1$) and $\beta_n$ denotes the condensation rate (the rate of cluster growth from size $n$ to $n + 1$). The concentration of monomers (single atoms), $c_1$, is constrained by the conservative condition of the total solute concentration $c_{\text{tot}}$:
\begin{equation}
    c_1 = c_{\text{tot}}- \sum_{n\geq2} {n c_n},
\end{equation}
With the quasi-equilibrium assumption at precipitate-matrix interfaces that determines the driving force of diffusion flux~\cite{perez2005gibbsthomson,perez2008implementation}, these rates present a relationship:
\begin{equation}
\frac{\beta_n}{\alpha_{n+1}} = c_1 \exp\left(-\frac{\Delta G_{n+1}-\Delta G_{n}-\Delta G_{1}}{k_B T}\right)
\end{equation}
where $\Delta G_n$ is the free energy of formation for a cluster of size $n$, composed of Mg and Zn atoms:
\begin{equation}
   \Delta G_n = \Delta H_n - T \Delta S_n
\end{equation}
Although the accurate description of multicomponent clusters in Al-Mg-Zn alloys requires treating $\Delta G_n$ as a function of both $n_{\text{Mg}}$ and $n_{\text{Zn}}$, to reduce complexity and ensure computational efficiency, we adopt a fixed Zn/Mg ratio of 2, with the rationale explained in the Discussion section. The enthalpy of formation $\Delta H_n$ is modeled using a fitting function~\cite{miyoshi2019temperaturedependent}:
\begin{equation}
    \Delta H_{n} = 
\begin{cases}
An-An^p & \text{for } n \leq q, \\
Bn+Cn^{1/2}+D & \text{for } n > q,
\end{cases}
\end{equation}
where $A=-0.969$ eV, $B=-0.250$ eV, $C=0.605$ eV, $D=-0.044$ eV, $p=0.952$, and $q=63$ are fitting parameters from the \ac{SA} simulation results. For the entropy term $\Delta S_n$, we only consider the configurational entropy contributions when solute atoms are in dilute solid solutions:
\begin{equation}
    \Delta S_n \approx  k_B \left[\frac{n}{3}\ln\left(c_{\text{tot}}^{\text{Mg}}\right)+\frac{2n}{3}\ln\left(c_{\text{tot}}^{\text{Zn}}\right)\right],
\end{equation}
where $c_{\text{tot}}^{\text{Mg}}$ and $c_{\text{tot}}^{\text{Zn}}$ are concentrations of each solute element present in the entire alloy, regardless of whether they are distributed as isolated solute atoms or within clusters. The condensation rate $\beta_n$ was explicitly modeled as~\cite{hirschfelder1974kinetics,trinkaus1983theory}:
\begin{equation}
\label{eq:beta_n}
    \beta_n = \frac{B_{11}B_{22}-{B_{12}}^2}{B_{11} \sin^2 \theta + B_{22} \cos^2 \theta},
\end{equation}
where angle $\theta$ denotes the nucleation and growth direction in the $(n_\text{Mg}, n_\text{Zn})$ space.  With Zn/Mg ratio of 2, $\theta = \arctan 2$. Here $B_{ij}$ in Eq.~\ref{eq:beta_n} is the element of matrix $\pmb B$, which that characterizes the diffusion-limited condensation process for the critical cluster~\cite{clouet2009modeling}:
\begin{equation}
    \pmb B= \frac{4\pi R_{n}}{\Omega} 
    \begin{pmatrix}
     { c_{1}^{\text{Mg}} D^\text{Mg}} & 0 \\
    0 & c_{1}^{\text{Zn}} D^\text{Zn}
    \end{pmatrix},
\end{equation}
where $R_{n}$ is the radius of a solute cluster containing $n$ atoms by assuming it is in a spherical shape, $\Omega$ is the Wigner-Seitz cell volume (the volume of the single atomic site). Both $R_{n}$ and $\Omega$ were calculated based on an \ac{FCC} lattice with a lattice constant of 4.046\r{A}. $c_{1}^{\text{Mg}}$ and $c_{1}^{\text{Zn}}$ are the concentration of single solute Mg and Zn, respectively. $D^\text{Mg}$ and $D^\text{Zn}$ are the solute diffusion parameters of Mg and Zn in Al:
\begin{equation}
     D^{\text{X}}= D^{\text{X}}_0 \exp \left(-{Q^{X}}/{k_B T}\right)  \frac{c^{\text{vac,res}}_{\text{dyn}} }{c^{\text{vac,Al}}_{\text{eq}} },
\end{equation}
Wu \textit{et al.}~\cite{wu2016highthroughput} provided first-principles solute diffusion constants, $D^{\text{X}}_0$ for Mg of $9.4182 \times10^{-2}$ cm$^2$/s and Zn of $1.0614\times10^{-1}$ cm$^2$/s, and solute diffusion activation energy (including both vacancy formation and migration energies), $Q^{X}$, for Mg and Zn in Al matrix of $1.2467$ eV and $1.1812$ eV, respectively. $c^{\text{vac,res}}_{\text{dyn}}$ is calculated based on Eq.~\ref{eq:vac_dyn_res2} using parameters $\tilde E_{\text{bind}}^{\text{clu}}$ obtained from \ac{KMC} and absorbing Markov chain models. $c^{\text{vac,Al}}_{\text{eq}}$ is the equilibrium vacancy concentration:
\begin{equation}
     c^{\text{vac,Al}}_{\text{eq}}=c_0 \exp\left(\frac{S^{\text{vac,Al}}}{k_B}\right)\exp\left(-\frac{H^{\text{vac,Al}}}{k_BT}\right)
\end{equation}
where we adopt the values from pure Al \cite{carling2003vacancy}: $c_0 = 1.67$ as the preexponential factor, $H^{\text{vac,Al}}= 0.66$ eV as the vacancy formation enthalpy and $S^{\text{vac,Al}}= 0.7k_B$ as the vacancy formation entropy.
\bibliography{reference}

\section*{Acknowledgements}
This research was supported by NSF grant \#1905421, with computing resources provided by GM Global Technical Center and Stampede3 at TACC (ACCESS allocation DMR190035, NSF grants \#2138259, \#2138286, \#2138307, \#2137603, \#2138296). The authors thank Prof. Dane Morgan from the University of Wisconsin-Madison for insightful discussions.

\section*{Author contributions statement}
Z.X. and L.H. performed the high-throughput first-principles simulations. Z.X. and L.Q. constructed the models and analyzed the results. L.H., A.M., and L.Q. conceptualized and supervised the research project. All authors analyzed the results and prepared the manuscript together.

\section*{Competing interests}
The authors declare no competing interests.
% To include, in this order: \textbf{Accession codes}(where applicable); \textbf{Competing interests}(mandatory statement). 

% The corresponding author is responsible for submitting a \href{http://www.nature.com/srep/policies/index.html#competing}{competing interests statement} on behalf of all authors of the paper. This statement must be included in the submitted article file.

\end{document}

\begin{acronym}
    \acro{APT}{atom probe tomography}
    \acro{ATAT}{Alloy Theoretic Automated Toolkit}
    \acro{BEP}{Br\text{\o}nsted-Evans-Polanyi}
    \acro{CI-NEB}{climbing image nudged elastic band}
    \acro{CMC}{canonical Monte Carlo}
    \acro{CNT}{classical nucleation theory}
    \acro{DD}{dislocation dynamics}
    \acro{DFT}{density function theory}
    \acro{ECI}{effective cluster interaction}
    \acro{EDS}{energy dispersive X-ray spectroscopy}
    \acro{FCC}{face-centered cubic}
    \acro{FEA}{finite element analysis}
    \acro{GGA}{generalized gradient approximation}
    \acro{GP}{Guinier-Preston}
    \acro{GB}{grain boundary}
    \acrodefplural{GB}{grain boundaries}
    \acro{KMC}{kinetic Monte Carlo}
    \acro{KRA}{kinetically resolved activation}
    \acro{LDA}{local density approximation}
    \acro{MC}{Monte Carlo}
    \acro{MD}{molecular dynamics}
    \acro{MEP}{minimum energy path}
    \acro{NEB}{nudged elastic band}
    \acro{PAW}{projector-augmented wave potentials}
    \acro{PBE}{Perdew-Burke-Ernzerhof}
    \acro{PBEsol}{PBE for solids}
    \acro{PCA}{principal component analysis}
    \acro{PEL}{potential energy landscape}
    \acro{PF}{phase-filed}
    \acro{PPM}{parts per million}
    \acro{PW91}{Perdew-Wang}
    \acro{SRC}{solute-rich clusters}
    \acro{SQS}{special quasi-random structure}
    \acro{SSSS}{supersaturated solid solution}
    \acro{STEM}{Scanning transmission electron microscopy}
    \acro{TEM}{transmission electron microscopy}
    \acro{VASP}{Vienna \textit{Ab Initio} Simulation Package}
    \acro{VTST}{Transition States Tools}
    \acro{RMSE}{root-mean-square error}
    \acro{R$^2$}{coefficient of determination}
    \acro{CD}{cluster dynamics}
    \acro{SA}{simulated annealing}
\end{acronym}
\flushbottom
\maketitle
\thispagestyle{empty}

\clearpage

\section{Supplementary Note \arabic{section}:}
\large{\sffamily{\bfseries{Surrogate models to predict vacancy migration barriers and driving forces}}}

We employ surrogate models that take the lattice occupation configuration as input to predict several key energetics: the total energy $E_\text{tot}$ of a given configuration within a supercell, the energy profile $E_{\text{MEP}}$ along the \ac{MEP} for vacancy migration as a function of the reaction coordinate, $x$, and critical energetic parameters related to vacancy migration, including the energy difference $E^{\text{diff}}_{\text{mig}}$ and the migration energy barrier $E^{\text{bond}}_{\text{mig}}$.

\begin{figure}[!htp]
    \centering
    \includegraphics[width=1\linewidth]{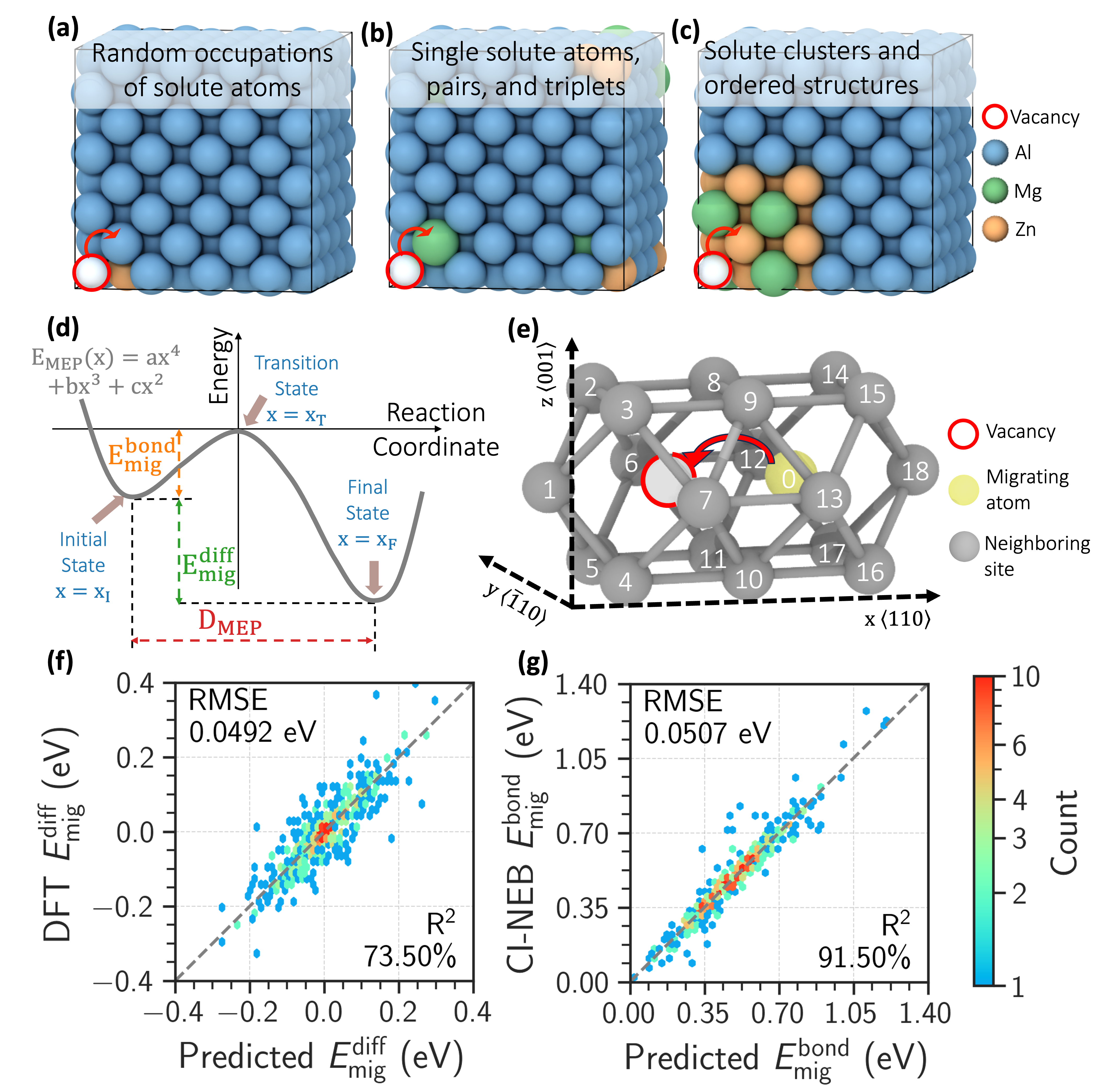}
    \caption{Surrogate models to predict MEP and the critical energetic parameters for vacancy migrations in Al-Mg-Zn alloy. (a)-(c): Schematic plots diagrams of 4$\times$4$\times$4 supercells used to calculate vacancy migration barriers, including three typical cases: supercells with one single solute atom embedded near the vacancies (a), supercells with random distributions of solutes (b), supercells with 2$\times$2$\times$2 ordered solute structures embedded on Al matrix. Red circles denote vacancies. Blue, orange, and green spheres represent Al, Zn, and Mg atoms, respectively. (d): Schematic plot of the MEP of vacancy migration showing $E^{\text{diff}}_{\text{mig}}$, $D_{\text{MEP}}$, and $E^{\text{bond}}_{\text{mig}}$ by using the quartic function to describe $E_{\text{MEP}}$ as a function of reaction coordinate $x$. (e): Illustrations of local lattice sites (1$^{st}$ neighboring lattice sites) related to the vacancy migration considered in the surrogate models, where the vacancy and the migration atom are in alignment with the $\langle110\rangle$ direction (x-axis). Each atom is labeled by an index. (f)-(g): Performance of surrogate models to predict $E^{\text{diff}}_{\text{mig}}$ (f) and $E^{\text{bond}}_{\text{mig}}$ (g) for the testing data set in Al-Mg-Zn alloys by comparisons with those from first-principles DFT+CI-NEB calculations. The root-mean-square error (RMSE) is denoted at the upper left, and the number at the bottom-right corner shows the coefficient of determination (R$^{2}$).}
    \label{fig:supplementary:surrogate_model}
\end{figure}

\paragraph{Surrogate models dataset} These surrogate models are trained using data from our previous first-principles study\cite{xi2022mechanism}, which computed 1250 \acp{MEP} using \ac{DFT} and \ac{CI-NEB} method~\cite{kresse1996efficiency,kresse1996efficient,blochl1994projector,perdew1996generalized,monkhorst1976special,henkelman2000improved,sheppard2008optimization}. This yielded a dataset of 2500 $E^{\text{diff}}_{\text{mig}}$ and $E^{\text{bond}}_{\text{mig}}$ data points, accounting for both forward and backward vacancy migrations, along with the corresponding initial and final supercell configurations and their total energies. 80$\%$ of the data points were selected as the training dataset, the remaining 20 $\%$ were set aside as a test set to assess the predictive performance of the model.

For the representativeness of the dataset, we conducted high-throughput \ac{DFT} calculations for Al-Mg binary alloys, Al-Zn binary alloys, and Al-Mg-Zn ternary alloys. All calculations used 4$\times$4$\times$4 supercells, constructed from the FCC Al unit cell, with 255 atoms and 1 vacancy. The supercells used to calculate the vacancy migration barriers are divided into three categories. The first category, as shown in Fig.~\ref{fig:supplementary:surrogate_model}a, consists of supercells with a single solute atom cluster (single atoms, pairs, or triplets) embedded in the neighboring site of the vacancy. These configurations address the effect of a single solute atom cluster on the vacancy migration barrier. The second category, as shown in Fig.~\ref{fig:supplementary:surrogate_model}b, consists of supercells that are randomly generated solid solution structures with different local concentrations of solute atoms around the vacancy. These structures simulate vacancy diffusion in the solid-solution states. The third category, as shown in Fig.~\ref{fig:supplementary:surrogate_model}c, consists of supercells with ordered cluster structures embedded in the Al matrix. These are designed to describe the vacancy moving within or near the precipitates. The ordered structures were chosen from proposed \ac{GP} zone precipitates in Al-Mg-Zn alloys~\cite{berg2001gpzones} and other ordered intermetallic structures (L1$_\text{0}$, L1$_\text{2}$, L1$_\text{0}^*$, W2, CH, and Z1) on a \ac{FCC} lattice\cite{zhuravlev2017phase}.

\paragraph{Surrogate models for total energy} The surrogate model to predict $E_\text{tot}$ of a lattice occupation configuration was guided by the cluster expansion~\cite{sanchez1984generalized,vanderven2001firstprinciples,zhang2016cluster}. The cutoff criteria we applied are that only clusters with no more than three sites were considered and that no two sites shall be further apart than the third nearest neighbor distance in each cluster. The input to the model was the vector to express the full lattice occupation configuration, $\vec{\sigma}_{\text{config}} $ with a length of $N \times M$, defined by the atom types at all lattice sites: 
\begin{equation}
    \vec \sigma_{\text{config}} = \left\{ \sigma_1^{(1)}, \sigma_1^{(2)},\cdots, \sigma_1^{(M)},\sigma_2^{(1)},\sigma_2^{(2)}, \cdots, \sigma_2^{(M)}, \cdots, \sigma_N^{(1)}, \sigma_N^{(2)}, \cdots, \sigma_N^{(M)} \right\},
\end{equation}
where $N$ is the total number of lattice sites in the configuration and $M$ is the number of chemical species in the alloy. Here, we denote the occupation of the chemical species $A$ on the site $i$ as $\sigma_i^{(A)}$. This means that for each site $i$, we apply a $M$-dimensional one-hot encoding vector $\vec \sigma_i = \{ \sigma_i^{(1)}, \cdots, \sigma_i^{(M)} \}$ to represent the occupation of the site. If $A$ atom is at the site $i$, $\sigma_i^{(A)}=1$ otherwise $\sigma_i^{(A)}=0$. Since each site can only be exactly one atom, we have $\sum_{l=1}^M {\sigma_i^{(l)}} = 1$. Using this one-hot encoding technique for categorical data, such as chemical species, can eliminate any quantitative relationship of variables compared to giving each site a scalar index. For a pair cluster of occupation $A$ and $B$ on sites $i$ and $j$, it can be described as $\sigma_{ij}^{(AB)} = \sigma_i^{(A)} \sigma_j^{(B)}$. We will have similar rules: $\sum_{l=1}^M \sum_{m=1}^M {\sigma_{ij}^{(lm)}} = 1$, which suggests that sites $i$ and $j$ can only be occupied once. This can be generalized to larger clusters. For example, a three-body cluster can be described as $\sigma_{ijk}^{(ABC)} = \sigma_i^{(A)} \sigma_j^{(B)} \sigma_k^{(C)}$.

$E_{\text{tot}}$ of the same system is then proportional to the number of lattices in the supercell. Hence, the number of each type of cluster is informative for the evaluations of $E_{\text{tot}}$ and other lattice-occupation-dependent properties. We use $\phi_{\alpha}$ to denote the number of a particular type of cluster in the supercell ($\alpha$ is the correlation function iterator). For instance, if we consider type $\alpha$ as the first nearest neighbor pairs of A-B, $\phi_{\alpha}  = \sum_{i,j} (N_1(i, j)\sigma_{ij}^{(\text{AB})})$, where $N_1(i, j)$ is an indicator function. $N_1(i, j) = 1$ if site $i$ and site $j$ is at first nearest neighbor distance, otherwise $N_1(i, j) = 0$.The complete feature vector $\vec{\phi}{\alpha}$ is formed by collecting all such $\phi{\alpha}$ values, representing the full configuration. Now the total energy can be expressed as:
\begin{equation}
\label{eq:supplementary:E_total}
    E_{\text{tot}}(\vec \sigma_{\text{config}}) = \sum_{\alpha}{\tilde{J}_{\alpha}} \phi_{\alpha},
\end{equation}
where $\tilde J_{\alpha}$ is the \ac{ECI} per atom, and $\alpha$ is a notation for a particular cluster type. The original formula of this equation is, $\tilde J_{\alpha} =J_{\alpha} {m_{\alpha}}$.
where $J_{\alpha}$ is the \ac{ECI} determined by the energy landscape of the material, and ${m_{\alpha}}$ is the multiplicity or the number of clusters $\alpha$. Usually, it is practical to combine ${m_{\alpha}}$ and $J_{\alpha}$ into one term, $\tilde J_{\alpha}$ as the effective \ac{ECI} per atom. 

\paragraph{Surrogate models for minimum energy path} The \ac{MEP} in Al-Mg-Zn alloys can be expressed as a quartic function\cite{xi2022mechanism}: 
\begin{equation}
\label{eq:supplementary:quartic}
E_{\text{MEP}}(x) = ax^4+bx^3+cx^2,
\end{equation}
where $E_{\text{MEP}}$ is the energy landscape of the \ac{MEP} as a function of the reaction coordinate $x$. Here, the coefficients ($a$, $b$, and $c$ in Eq.~\ref{eq:supplementary:quartic}) depend on the local lattice distortions introduced by different occupations on the sites near the vacancy and the migration atom. As shown in Fig.~\ref{fig:supplementary:surrogate_model}d, the energy landscape has a transition state ($x_\text{T}$) at the original point ($x_\text{T}=0$), and the positions of the initial state ($x_\text{I}$) and final state ($x_\text{F}$) at two local minima ($x_\text{I}=\frac{-3b-\sqrt{9b^2-32ac}}{8a}$, and $x_\text{F}=\frac{-3b+\sqrt{9b^2-32ac}}{8a}$). Thus, the critical \ac{MEP} properties can be calculated by the parameters of the coefficients: the vacancy migration barrier, $E^{\text{bond}}_{\text{mig}} =-E_{\text{MEP}}(x_{\text{I}})$, the energetic driving forces, $E^{\text{diff}}_{\text{mig}} = E_{\text{MEP}}(x_{\text{F}})-E_{\text{MEP}}(x_{\text{I}}) = b^2(x_{\text{F}}^3-x_{\text{I}}^3) = \frac{b}{256 a^3}\left(9 b^2-32 a c\right)^{3/2}$, the vacancy migration distance, $D_{\text{MEP}}=x_{\text{F}}-x_{\text{I}}=\frac{\sqrt{9 b^2-32 a c}}{4 a}$, and the summation of second derivatives at the initial ($K_{\text{I}}$) and final state ($K_{\text{F}}$), $K_{\text{sum}} = K_{\text{I}} + K_{\text{F}} = E_{\text{MEP}}''(x_{\text{I}})+E_{\text{MEP}}''(x_{\text{F}})$. 

To unify our energetic surrogate models and make sure accuracy at both short-range distance and long-range distance, we directly calculate the the energy difference along the \ac{MEP}, $E^{\text{diff}}_{\text{mig}} = E_{\text{MEP}}(x_{\text{F}}) - E_{\text{MEP}}(x_{\text{I}})$, where both $E_{\text{MEP}}(x_{\text{F}})$ and $E_{\text{MEP}}(x_{\text{I}})$ are predicted using the cluster expansion total energy surrogate models of Eq.~\ref{eq:supplementary:E_total}. Subsequently, certain properties of the MEP, such as $D_{\text{MEP}}$ and $K_{\text{sum}}$, can be characterized by the surrogate model from our previous work. Consequently, quartic coefficients can be expressed in the term of $E^{\text{diff}}_{\text{mig}}$, $D_{\text{MEP}}$ and $K_{\text{sum}}$ based on Eq.~\ref{eq:supplementary:quartic}: $a=\frac{D_{\text{MEP}}^2 K_{\text{sum}}}{4}$, $b=\frac{4 E^{\text{diff}}_{\text{mig}}}{D_{\text{MEP}}^3}$, and $c=\frac{144 {E^{\text{diff}}_{\text{mig}}}^2-D_{\text{MEP}}^{12} K_{\text{sum}}^2}{8 D_{\text{MEP}}^8 S}$. Finally, $E^{\text{bond}}_{\text{mig}}$ is formulated as:
\begin{equation}
\label{eq:supplementary:Ea}
E^{\text{bond}}_{\text{mig}} = \frac{\left(D_{\text{MEP}}^6 K_{\text{sum}}-12 {E^{\text{diff}}_{\text{mig}}}\right)^3 \left(4 {E^{\text{diff}}_{\text{mig}}}+D_{\text{MEP}}^6 K_{\text{sum}}\right)}{64 D_{\text{MEP}}^{18} K_{\text{sum}}^3}.
\end{equation}

For the prediction of $E^{\text{bond}}_{\text{mig}}$, the input was the type of the migrating atom and the type of all atoms on the $1^{\text{st}}$, $2^{\text{nd}}$, and $3^{\text{rd}}$ nearest-neighbor lattice sites relative to the vacancy/migration atom, $\vec \sigma_{\text{neighbor}}$. An illustration of the 1$^{st}$ nearest neighbor lattice sites can be found in Fig.~\ref{fig:supplementary:surrogate_model}e. Based on the chemical type of the migrating atom (Al, Mg, or Zn), the training data were divided into three different groups for \ac{MEP} predictions. Then we applied a regression for $D_{\text{MEP}}$ and $K_{\text{sum}}$ with the vector $\vec \sigma_{\text{neighbor}}$. We notice that $D_{\text{MEP}}$ remains consistent when a vacancy migrates from the initial state to the final state, as well as when jumping back. However, the second derivatives at the initial and final states are different. Hence, we applied different symmetry constraints on the correlation function iterators $\beta$ and $\gamma$, which are used in the cluster expansion methods to fit the values of $D_{\text{MEP}}$ and $K$. The regressions can be expressed as:
\begin{equation}
    D_{\text{MEP}}(\vec \sigma_{\text{neighbor}}) = \sum_{\beta}{\tilde J_{\beta}} \phi_{\beta},
\end{equation}
and
\begin{equation}
    K(\vec \sigma_{\text{neighbor}}) = \sum_{\gamma}{\tilde J_{\gamma}}  \phi_{\gamma},
\end{equation}
where $K$ can be either $K_{\text{I}}$ or $K_{\text{F}}$. 

For the regression of $D_{\text{MEP}}$, we employed the symmetry operations of the $mmm$ point group with a mirror symmetry plane perpendicular to the $\langle\bar{1}10\rangle$ (y-axis), a mirror symmetry plane perpendicular to the $\langle001\rangle$ (z-axis), and a mirror symmetry plane perpendicular to the $\langle110\rangle$ direction (x-axis). By applying these symmetry operations, lattice sites within the crystal structure excluding the migration atom site can be systematically divided into multiple distinct sets, i.e., \{1, 18\}, \{2, 3, 4, 5, 14, 15, 16, 17\}, \{6, 7, 12, 13\}, and \{8, 9, 10, 11\} for the 1$^{\text{st}}$ nearest neighbor lattice sites shown in Fig.~\ref{fig:supplementary:surrogate_model}e. Similarly, for the regression of $K_{\text{I}}$ and $K_{\text{F}}$, we employed the symmetry operations of the $mm2$ point group with a mirror symmetry plane perpendicular to the $\langle\bar{1}10\rangle$ (y-axis), a mirror plane perpendicular to the $\langle001\rangle$ (z-axis), and a 2-fold rotation axis along the $\langle110\rangle$ direction (x-axis). The lattice sites can be grouped as \{1\}, \{2, 3, 4, 5\}, \{6, 7\}, \{8, 9, 10, 11\}, \{12, 13\}, \{14, 15, 16, 17\}, and \{18\} for the 1$^{\text{st}}$ nearest neighbor lattice sites. 

Lattice sites belonging to the same set share similar characteristics and therefore can be averaged. In this case, if we consider the first nearest neighbor pairs of Mg-Zn as an instance, there will be one set of parameters to describe them, $\left\{ \phi_{\beta_1} =\sum_{i,j} ({C_1(i,j)N_1(i, j)\sigma_{ij}^{(\text{MgZn})}}), \cdots, \phi_{\beta_k}=\sum_{i,j} ({C_k(i,j)N_1(i, j)\sigma_{ij}^{(\text{MgZn})}}) \right\}$. Here, $C_k(i,j)$ are also indicator functions (either 0 or 1) to show whether site $i$ and site $j$ are from particular groups. Each component in the set represents one type of the first nearest Mg-Zn pair that has a particular relative position to the vacancy/migration atom.  When we extended this method to other 3-atoms clusters within $3^{\text{rd}}$ nearest neighboring distance of the vacancy/migration atom, we obtained the combined feature vectors, $\vec \phi_{\beta}$ and $\vec \phi_{\gamma}$, that describe the local environment of a vacancy migration event with the symmetry of $mmm$ or $mm2$. 

The combined feature vectors, $\vec \phi_{\beta}$ and $\vec \phi_{\gamma}$, which encode lattice occupations at the lattice site, have dimensionalities of 1401 and 711 when constructed using the point group symmetry operations $mm2$ and $mmm$, respectively. The dimensionalities of vectors are considerably large due to the utilization of one-hot encoding for labeling different chemical species. This implies that the vector contains a significant number of redundant digits, which can be effectively reduced using the \ac{PCA} method. After the dimensionality reduction, linear regression with the $L_2$ regularization was applied to the training data for the prediction of $D_{\text{MEP}}$ and $K$. As a result, $E^{\text{bond}}_{\text{mig}}$ can be obtained by applying Eq.~\ref{eq:supplementary:Ea}.

\paragraph{Surrogate models performance} The accuracy and performance of the surrogate models were validated by comparison with \ac{DFT}+\ac{CI-NEB} calculations. First, predictions of $E^{\text{diff}}_{\text{mig}}$ were assessed by calculating the difference in the predicted $E_{\text{tot}}$ values between the initial and final states of given lattice occupation configurations. As shown in Fig.~\ref{fig:supplementary:surrogate_model}f, the predicted $E^{\text{diff}}_{\text{mig}}$ (X-axis) closely aligns with the values derived from \ac{DFT} calculations (Y-axis), achieving a low \ac{RMSE} of 0.0492 eV and a \ac{R$^2$} score of 73.50$\%$. 

The performance of the surrogate model in predicting \ac{MEP} energy barriers ($E^{\text{bond}}_{\text{mig}}$) was also evaluated, as illustrated in Fig.~\ref{fig:supplementary:surrogate_model}g. The comparison between predicted $E^{\text{bond}}_{\text{mig}}$ (X-axis) and \ac{CI-NEB} results (Y-axis) yielded an \ac{RMSE} of 0.0507 eV and a high \ac{R$^2$} score of 91.50$\%$, indicating strong predictive accuracy for $E^{\text{bond}}_{\text{mig}}$. Small RMSE values and higher R$^2$ scores for both models reflect the capability of surrogate models to capture these energetics effectively.

\clearpage

\section{Supplementary Note \arabic{section}:}
\large{\sffamily{\bfseries{Simulated annealing simulations for solute clusters}}}
The \ac{SA} heuristic shares similarities with \ac{CMC} simulations but differs in that it operates with a decreasing temperature throughout the simulation steps to locate the global minimum configuration. Despite this gradual temperature reduction, \ac{SA} can still be employed to quantify the system's equilibrium under a fixed volume of the simulation supercell and a predefined number of each chemical element within the system. In this algorithm, at each step, the energy $E(\vec{\sigma})$ of the atomic configuration $\vec{\sigma}$ was evaluated using a surrogate model. Swaps between atom pairs of different species were proposed, excluding swaps between identical atoms to improve efficiency. Although the temperature changes during the simulation, a quasi-equilibrium assumption is maintained at each step. The acceptance of a proposed new state $\sigma_j$ from the current state $\sigma_i$ followed the Metropolis criterion based on the Boltzmann distribution:
\begin{equation}
p_{ij} =
\begin{cases}
1 & \text{for } \Delta E \leq 0, \\
\exp\left(-\frac{\Delta E}{k_B T} \right) & \text{for } \Delta E > 0,
\end{cases}
\end{equation}
where $\Delta E = E(\sigma_j) - E(\sigma_i)$. This algorithm allows for occasional increases in total energy, which prevents the system from becoming trapped in local minima. As the temperature decreases, the probability of escaping local minima gradually diminishes until it approaches 0 K, enabling the identification of the global minimum configuration.

Mg and Zn atoms were randomly distributed according to specified values of $n_{\text{Mg}}$ and $n_{\text{Zn}}$.  To avoid the effects of solute interaction in the initial configurations to the formation energy, we do not directly use the energy difference between the initial and final states for comparison. Here, we ignore the effects of temperatures and external pressure to obtain the minimum energy of a solute cluster (${H}_\text{form} \approx {E}_\text{form}$), and the average formation enthalpy $\bar{H}_\text{form}$ of a cluster is described as:
\begin{equation}
     \bar{H}_\text{form}=\frac{E_\text{final}  - n_{\text{Mg}} \mu_{\text{Mg}} -  n_{\text{Zn}}\mu_{\text{Zn}} - \left({N-n_{\text{Mg}}-n_{\text{Zn}}} \right)E_\text{Al}}{n_{\text{Mg}}+n_{\text{Zn}}}.
\end{equation}
Here, $E_\text{final}$ is the final energy obtained from the \ac{SA} simulations. $E_\text{Al}$ is the energy of a single Al atom in a pure Al crystal. $\mu_{\text{Mg}}$ and $\mu_{\text{Zn}}$ are chemical potentials of solute Mg and Zn, respectively, and can be expressed as:
\begin{equation}
   \mu_{\text{X}} = E \left({\text{Al}}_{N-1}\text{X}_{1} \right) - (N-1)E_\text{Al},
\end{equation}
where $E({\text{Al}}_{N-1}\text{X})$ is the energy of an Al supercell containing a single solute atom X. \Ac{SA} was implemented using a gradually decreasing temperature schedule:
\begin{equation}
T_n = T_0 \left(1 - \frac{\alpha}{n_{\text{max}}} \right)^n,
\end{equation}
where $\alpha$ is the decay factor and is chosen to be $3$, and $n_{max}$ is the total simulation step of $10^{9}$. When we start from $T_0=800$ K, by the final step, $T_n \approx 39.8$~K, approximating a zero-temperature search. 

\begin{figure}[!htbp]
    \centering
    \includegraphics[width=1.0\linewidth]{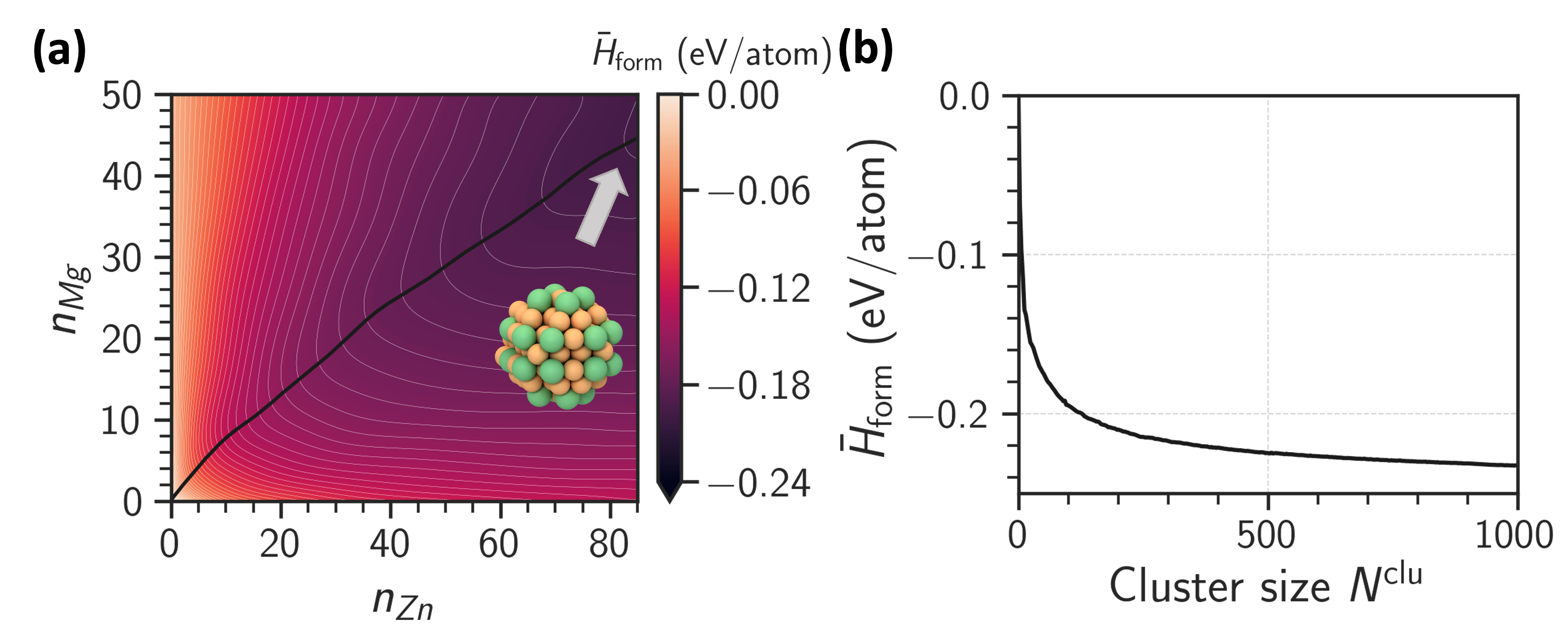}
    \caption{Formation enthalpy per atom, $\bar H_\text{form}$, of solute cluster obtained from SA. (a) Contour plot of $\bar H_\text{form}$ of solute clusters with respect to $n_{\text{Mg}}$ and $n_{\text{Zn}}$ (the numbers of Mg and Zn atoms in the cluster). The black line indicates the minimum energy pathway for energy dropping. The formed cluster is also presented, where orange spheres denote Zn atoms and green spheres are Mg atoms. (b) Evolution of $\bar H_\text{form}$ of solute cluster (with fixed Zn/Mg ratio of 2) with respect to cluster size $N^{\text{clu}}$.}
    \label{fig:supplementary:formation_energies}
\end{figure}

We first performed \ac{SA} simulation runs with varying solute compositions, where $n_\text{Mg}$ ranged from $0$ to $50$ and $n_\text{Zn}$ from $0$ to $85$. The results are summarized in the contour plot shown in Fig.~\ref{fig:supplementary:formation_energies}a, which maps $\bar{H}_\text{form}$ over the $(n_\text{Mg}, n_\text{Zn})$ space. The black line highlights the minimum-energy pathway, highlighting the most energetically favorable routes for cluster growth. Clusters along this pathway exhibit a Zn/Mg atomic ratio of approximately 2. 

Due to computational limits, results from the contour plot of Fig.~\ref{fig:supplementary:formation_energies}a were obtained from simulations of a $10 \times 10 \times 10$ FCC supercell. The largest cluster size explored there was $\sim$120 atoms. However, the clusters observed experimentally can reach several nanometers in diameter~\cite{chatterjee2022situ,geng2022earlystage}. To extend the analysis, we fixed the Zn/Mg ratio at 2 and performed additional simulations up to 999 solute atoms in a larger $17 \times 17 \times 17$ supercell. The resulting size dependence of $\bar{H}_\text{form}$ is shown in Fig.~\ref{fig:supplementary:formation_energies}b. We observed a rapid decrease in $\bar{H}_\text{form}$ for small clusters, followed by saturation near $-0.20$ eV/atom, indicating reduced sensitivity to further size increases.

\clearpage
\section{Supplementary Note \arabic{section}:}
\large{\sffamily{\bfseries{Vacancy escape time comparison for clusters from simulated annealing and kinetic Monte Carlo}}}

\begin{figure}[!htp]
    \centering
    \includegraphics[width=1\linewidth]{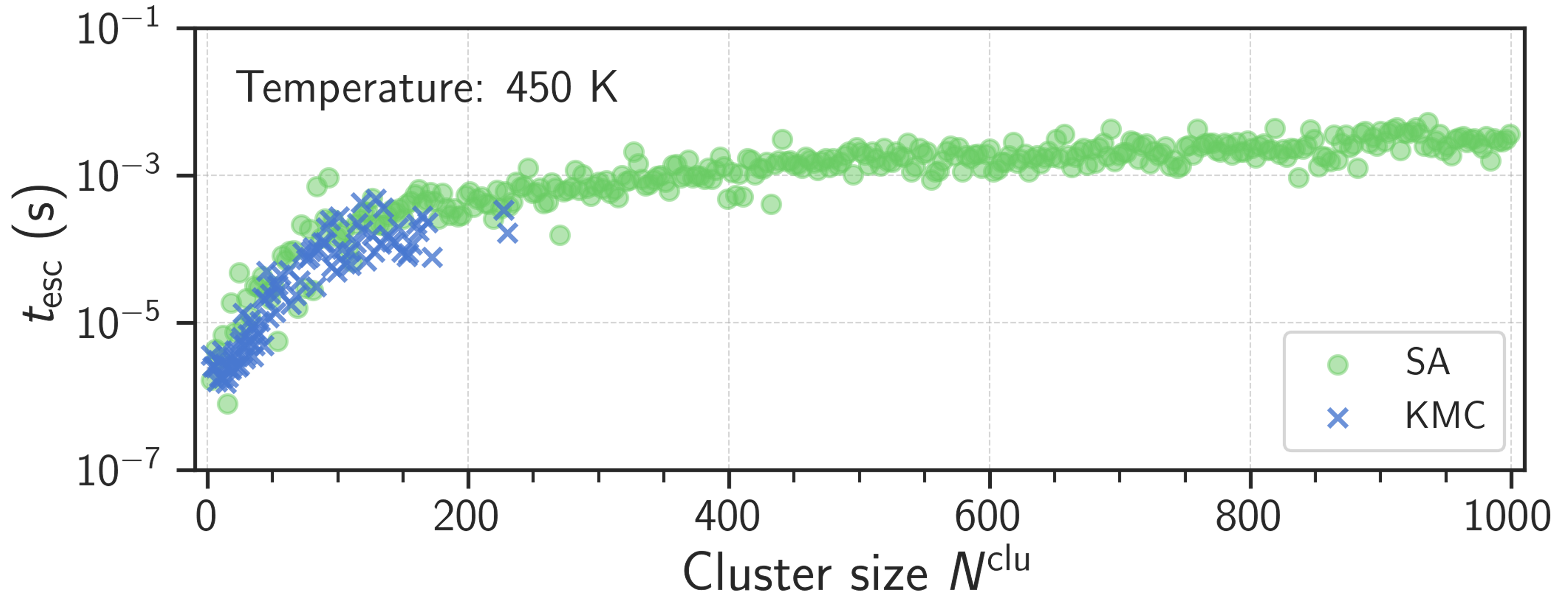}
    \caption{Comparison of the vacancy escape time from the absorbing Markov chain mode for clusters from simulated annealing and kinetic Monte Carlo. Green circles represent the clusters from SA simulations, and blue crosses denote clusters from KMC simulations.}
    \label{fig:supplementary:escape_time_comparison}
\end{figure}

To validate the applicability of \ac{SA}-generated cluster structures in describing the size-dependent effective vacancy binding energies $\tilde E_{\text{bind}}^{\text{clu}}$  we compare their predicted vacancy escape times $t_\text{esc}$ to those obtained from \ac{KMC} simulations conducted at 450 K. The \ac{KMC} simulation was performed in a $30 \times 30 \times 30$ \ac{FCC} supercell of Al containing 2.86 at.\% Mg and 2.38 at.\% Zn, with initial configurations equilibrated at 800 K via \ac{CMC} simulations. Fig.~\ref{fig:supplementary:escape_time_comparison} shows the vacancy escape time, $t_{\text{esc}}$, calculated from an absorbing Markov chain model for clusters derived from both SA and KMC simulations. The two sets of data show good agreement, both \ac{SA} and \ac{KMC} exhibit similar escape time magnitudes and scaling trends. These results at higher 450~K confirm that \ac{SA}-generated cluster configurations provide a reliable and consistent structural basis for modeling low-temperature (lower than 450~K) vacancy kinetics. These findings support the validity of using \ac{SA}-generated cluster structures as representative inputs in the subsequent lower-temperature (300~K) \ac{CD} modeling scenarios.

\clearpage

\section{Supplementary Note \arabic{section}:}
\large{\sffamily{\bfseries{Snapshots for as-quenched clusters obtained from different cooling conditions}}}

\begin{figure}[!htp]
    \centering
    \includegraphics[width=1\linewidth]{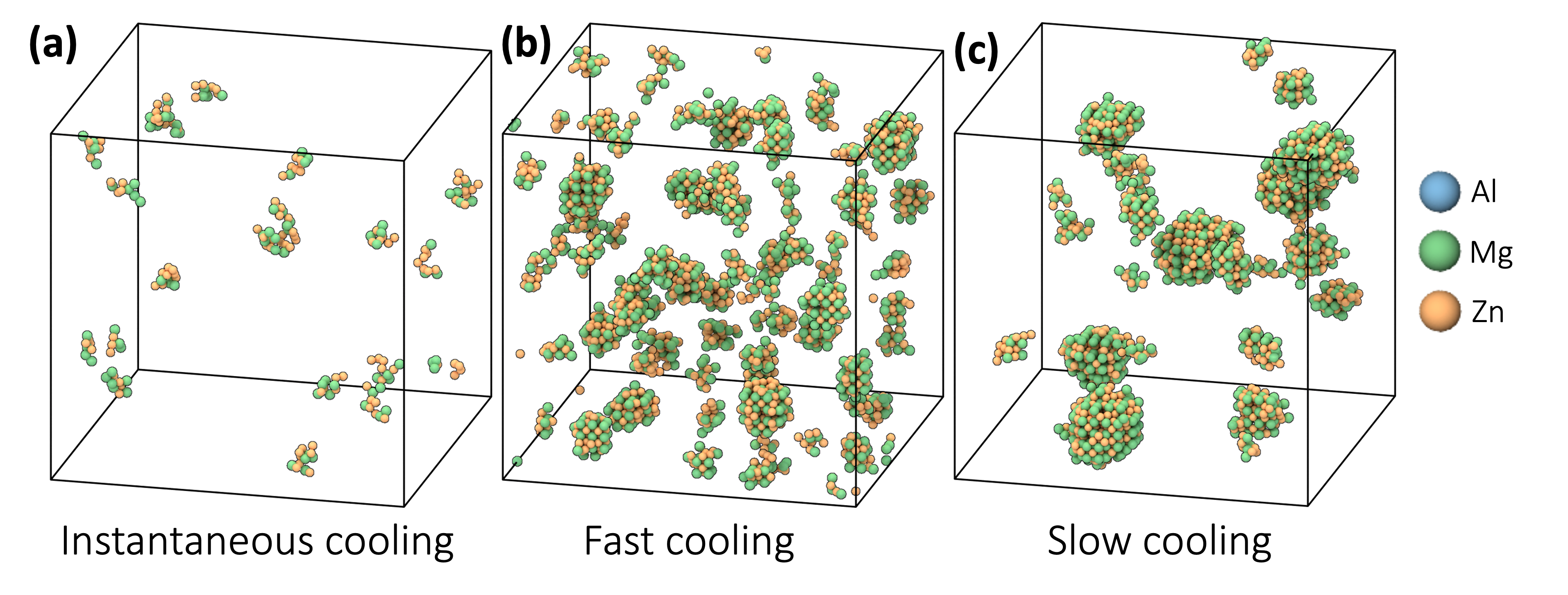}
    \caption{Snapshots of as-quenched clusters obtained from KMC simulations under different cooling conditions: (a) instantaneous cooling at the first step, (b) fast cooling at 10 s, and (c) slow cooling at 10 s. Only clusters containing more than 10 atoms are shown.}
    \label{fig:supplementary:aq_snapshots}
\end{figure}

Fig.~\ref{fig:supplementary:aq_snapshots} presents representative snapshots of as-quenched clusters obtained via \ac{KMC} simulations under different cooling conditions. The simulations were conducted in a $30 \times 30 \times 30$ \ac{FCC} supercell of Al-2.86 at.\%Mg-2.38 at.\%Zn. Initial configurations were equilibrated at 800 K using \ac{CMC}.  During quenching, the temperature was updated each timestep following predefined cooling profiles. These configurations provide initial inputs for subsequent \ac{CD} simulations. The snapshot corresponding to the infinitely fast (instantaneous) cooling rate (Fig.~\ref{fig:supplementary:aq_snapshots}a) was obtained immediately after the quench process, reflecting the supersaturated solid solution state. The snapshots for fast cooling (Fig.~\ref{fig:supplementary:aq_snapshots}b) and slow cooling (Fig.~\ref{fig:supplementary:aq_snapshots}c) conditions were recorded 10 seconds after beginning the quenching. 

\clearpage

\renewcommand*{\refname}{Supplementary References}

\bibliography{reference.bib}